\documentclass[letterpaper,11pt]{article}

\usepackage{diagbox}
\usepackage{makecell}
\usepackage{booktabs}
\usepackage{tikz,tikz-3dplot}
\usepackage{amsmath}
\usepackage{bbm}
\usepackage{amsfonts}
\usepackage{amssymb}
\usepackage{amsthm}
\usepackage{url,ifthen}
\usepackage[shortlabels]{enumitem}
\usepackage{srcltx}
\usepackage{dsfont}
\usepackage{multirow}
\usepackage{boxedminipage}
\usepackage[margin=1.1in]{geometry}
\usepackage{nicefrac}
\usepackage{xspace}
\usepackage{graphicx}
\usepackage{color}
\usepackage{subfigure}
\usepackage{colortbl}
\usepackage{setspace}
\usepackage{pgfplots}
\pgfplotsset{compat=1.12}
\usepackage{algorithm,algorithmic}
\usepackage[algo2e]{algorithm2e}

\usepackage{xcolor}
\definecolor{DarkGreen}{rgb}{0.1,0.5,0.1}
\definecolor{DarkRed}{rgb}{0.5,0.1,0.1}
\definecolor{DarkBlue}{rgb}{0.1,0.1,0.5}
\definecolor{Gray}{rgb}{0.2,0.2,0.2}

\definecolor{c1}{RGB}{38, 70, 83}
\definecolor{c2}{RGB}{42, 157, 143}
\definecolor{c3}{RGB}{233, 196, 106}
\definecolor{c5}{RGB}{231, 111, 81}
\definecolor{c4}{RGB}{244, 162, 97}

\definecolor{c1}{RGB}{38, 70, 83}
\definecolor{c2}{RGB}{42, 157, 143}
\definecolor{c3}{RGB}{233, 196, 106}
\definecolor{c5}{RGB}{231, 111, 81}
\definecolor{c4}{RGB}{244, 162, 97}

\usepackage{listings}
\lstdefinestyle{mystyle}{
    commentstyle=\color{DarkBlue},
    keywordstyle=\color{DarkRed},
    numberstyle=\tiny\color{Gray},
    stringstyle=\color{DarkGreen},
    basicstyle=\footnotesize,
    breakatwhitespace=false,         
    breaklines=true,                 
    captionpos=b,                    
    keepspaces=true,                 
    numbers=left,                    
    numbersep=5pt,                  
    showspaces=false,                
    showstringspaces=false,
    showtabs=false,                  
    tabsize=2
}
\usepackage[small]{caption}
\usepackage[pdftex]{hyperref}
\hypersetup{
    unicode=false,          
    pdftoolbar=true,        
    pdfmenubar=true,        
    pdffitwindow=false,      
    pdfnewwindow=true,      
    colorlinks=true,       
    linkcolor=DarkBlue,          
    citecolor=DarkGreen,        
    filecolor=DarkRed,      
    urlcolor=DarkBlue,          
    pdftitle={},
    pdfauthor={},
}

\def\draft{1}

\def\submit{0}

\ifnum\draft=1 
    \def\ShowAuthNotes{1}
\else
    \def\ShowAuthNotes{0}
\fi

\ifnum\submit=1
\newcommand{\forsubmit}[1]{#1}
\newcommand{\forreals}[1]{}
\else
\newcommand{\forreals}[1]{#1}
\newcommand{\forsubmit}[1]{}
\fi

\ifnum\ShowAuthNotes=1
\newcommand{\authnote}[2]{{ \footnotesize \bf{\color{DarkRed}[#1's Note:
{\color{DarkBlue}#2}]}}}
\else
\newcommand{\authnote}[2]{}
\fi

\newtheorem{theorem}{Theorem}[section]

\newtheorem{assumption}[theorem]{Assumption}

\newtheorem*{definition*}{Definition}

\theoremstyle{definition}

\newtheoremstyle{example_contd}
{\topsep} {\topsep}%
{}
{}
{\bfseries}
{.}
{1em}
{\thmname{#1} \thmnumber{ #2}\thmnote{#3} (continued)}

\theoremstyle{example_contd}

\newcommand{\chapterref}[1]{\hyperref[ch:#1]{Chapter~\ref{ch:#1}}}

\newcommand{\claimref}[1]{\hyperref[claim:#1]{Claim~\ref{claim:#1}}}

\newcommand{\corollaryref}[1]{\hyperref[cor:#1]{Corollary~\ref{cor:#1}}}

\newcommand{\definitionref}[1]{\hyperref[def:#1]{Definition~\ref{def:#1}}}

\newcommand{\equationref}[1]{\hyperref[eq:#1]{Equation~\ref{eq:#1}}}

\newcommand{\factref}[1]{\hyperref[fact:#1]{Fact~\ref{fact:#1}}}
\newcommand{\figurelabel}[1]{\label{fig:#1}}
\newcommand{\figureref}[1]{\hyperref[fig:#1]{Figure~\ref{fig:#1}}}
\newcommand{\tablelabel}[1]{\label{tab:#1}}
\newcommand{\tableref}[1]{\hyperref[tab:#1]{Table~\ref{tab:#1}}}

\newcommand{\itemref}[1]{\hyperref[item:#1]{Item~(\ref{item:#1})}}

\newcommand{\lemmaref}[1]{\hyperref[lem:#1]{Lemma~\ref{lem:#1}}}

\newcommand{\propref}[1]{\hyperref[prop:#1]{Proposition~\ref{prop:#1}}}

\newcommand{\propositionref}[1]{\hyperref[prop:#1]{Proposition~\ref{prop:#1}}}

\newcommand{\remarkref}[1]{\hyperref[rem:#1]{Remark~\ref{rem:#1}}}
\newcommand{\sectionlabel}[1]{\label{sec:#1}}
\newcommand{\sectionref}[1]{\hyperref[sec:#1]{Section~\ref{sec:#1}}}

\newcommand{\theoremref}[1]{\hyperref[thm:#1]{Theorem~\ref{thm:#1}}}

\usepackage[charter]{mathdesign}

\usepackage{microtype}

\newcommand{\Esymb}{\mathbb{E}}
\newcommand{\Psymb}{\mathbb{P}}

\DeclareMathOperator*{\E}{\Esymb}

\DeclareMathOperator*{\ProbOp}{\Psymb r}
\renewcommand{\Pr}{\ProbOp}

\renewcommand{\hat}{\widehat}

\newcommand{\defeq}{\stackrel{\small \mathrm{def}}{=}}
\renewcommand{\leq}{\leqslant}

\newcommand{\norm}[1]{\lVert#1\rVert_2}

\usepackage{bm}

\newcommand{\indep}{\!\perp\!\!\!\perp}
\newcommand{\notindep}{\not\!\perp\!\!\!\perp }

\newcommand{\ignore}[1]{}

\renewcommand{\epsilon}{\varepsilon}

\newcommand{\remove}[1]{}

\newcommand{\tmh}{\tau_{mod \rightarrow high}}
\newcommand{\btmh}{\overline{\tau}_{mod \rightarrow high}}
\newcommand{\tml}{\tau_{low \rightarrow mod}}
\newcommand{\btml}{\overline{\tau}_{low \rightarrow mod}}

\usepackage{fontawesome}

\usepackage{caption}
\usepackage{xcolor}
\usepackage{cite}
\usepackage{url}
\usepackage{tabularx} 

\usepackage{xspace}
\usepackage[bottom]{footmisc}
\usepackage{soul}

\usepackage[square]{natbib}

\newcommand{\ts}{t^*}

\newcommand{\Yhat}{\hat{Y}}

\title{Difficult Lessons on Social Prediction \\ 
from Wisconsin Public Schools} 

\date{\today}
\author{Juan C. Perdomo\thanks{University of California, Berkeley}\and Tolani Britton\footnotemark[1]\and Moritz Hardt\thanks{Max Planck Institute for Intelligent Systems, T\"ubingen, Germany, and T\"ubingen AI Center}\and Rediet Abebe\thanks{Harvard Society of Fellows}}

\begin{document}

\maketitle

\begin{abstract}
Early warning systems (EWS) are predictive tools at the center of recent efforts to improve graduation rates in public schools across the United States. These systems assist in targeting interventions to individual students by predicting which students are at risk of dropping out. Despite significant investments in their widespread adoption, there remain large gaps in our understanding of the efficacy of EWS, and the role of statistical risk scores in education. 

In this work, we draw on nearly a decade's worth of data and models from a system used throughout the State of Wisconsin to provide the first large-scale evaluation of the long-term impact of EWS on graduation outcomes. We present empirical evidence that the prediction system accurately sorts Wisconsin's students by their dropout risk. We also find that it may have caused a single-digit percentage increase in graduation rates, though our empirical analyses cannot reliably rule out that there has been no positive treatment effect. 

Going beyond a retrospective evaluation of DEWS, we draw attention to a central question at the heart of the use of early warning systems: Are individual risk scores necessary for effectively targeting interventions? We propose a simple mechanism that only uses information about students' environments---such as their communities, schools, and districts---and argue that this mechanism can target interventions just as efficiently as the individual risk score-based mechanism. Our argument holds even if individual risk predictions are highly accurate and effective interventions exist. In addition to motivating this simple targeting mechanism, our work provides a novel empirical backbone for the robust qualitative understanding among education researchers that dropout is structurally determined. 
Combined, our insights call into question the marginal value of individual predictions in settings where outcomes are driven by high levels of inequality.

\end{abstract}

\section{Introduction}

A class of automated risk prediction tools known as early warning systems (EWS) has recently become part of the de facto approach toward improving low high school graduation rates across the United States. Spurred by increased data collection efforts resulting from the 2001 No Child Left Behind Act, EWS were part of the public response to the so-called ``dropout crisis” of the early 2000s. Following initial pilot programs, EWS quickly boomed in use nationwide \citep{allensworth2007matters,balfanz2007preventing}. By 2015, over half of US public schools had implemented some version of an EWS, according to a survey by the Department of Education \citep{survey}. 

Early warning systems aim to increase graduation rates by improving how schools allocate educational resources to their students. These tools typically use data about students, schools, and districts to predict the ``individual probability'' that each student will drop out of high school. The scores are then examined by educators and school administrators to efficiently target interventions to students at high risk of not completing their high school degrees \citep{mac2019efficacy}.

Despite their surge in popularity and significant financial investment by education departments across the country, we lack conclusive evidence regarding these predictive systems' efficacy in reducing dropout rates. Challenges in implementing such empirical investigations and time lags in measuring their impact pose key barriers. As a result, existing studies on EWS focus on short-term impacts, are conducted on relatively small sample sizes, and present inconclusive results on the bottom-line effect on graduation rates \citep{balfanz2019early}.

In this work, we use a decade's worth of data from the Dropout Early Warning System (DEWS)---designed and implemented by the Wisconsin Department of Public Instruction (DPI)---to provide the first large-scale evaluation of the long-term impacts of these programs. Using an order of magnitude more data than previous studies on EWS, we show that the system accurately sorts students by their dropout risk and may have improved graduation rates. However, these effect estimates are noisy and we cannot rule out the possibility that the system has had no impact on student outcomes. 

Taking a more holistic view of the central objective of EWS, we propose an alternative resource allocation mechanism that relies on information about students' environments rather than individual risk scores. Grounding our arguments in data provided by the DPI, we illustrate how this alternative mechanism is significantly simpler to implement and justify to program participants while remaining just as effective in targeting interventions to students. This result comparing our environmental feature-based mechanism to individual risk score-based ones hold regardless of the accuracy of individual predictions and for the entire range of plausible intervention effects.

Beginning with the DEWS evaluation, we establish that the individualized predictions made by the program accurately sort students according to their true risk of dropping out of school. Further, DEWS provides a more accurate assessment of dropout risk for students from historically marginalized groups relative to their socioeconomically advantaged peers. However, these differences in predictive performance may have systematically de-prioritized marginalized students from receiving additional educational resources.

Apart from evaluating the system's accuracy, we investigate the causal impact of educational interventions assigned on the basis of DEWS predictions. In other words, we ask: Does DEWS usage improve high school graduation rates?
Leveraging the fact that DEWS risk categories are determined by thresholding a continuous score, we use a regression discontinuity design to estimate this treatment effect. Assigning students to the high-priority group may have led to a 5\% increase in their likelihood of on-time graduation.
However, this point estimate comes with non-trivial uncertainty: The 95\% confidence interval ranges from -2\% to 12\%. We cannot, therefore, reliably rule out the possibility that DEWS has had no positive treatment effect on graduation outcomes, even when we restrict our analyses to schools that actively use the system. 

A possible explanation regarding the noisy, single-digit treatment effect estimate is that schools lack sufficient guidance and resources to translate high-risk predictions into effective interventions. DPI has expanded efforts to support the use of the DEWS program throughout the state. Amongst other initiatives, they have collaborated with organizations such as the Wisconsin Response to Intervention Center to organize on-site DEWS training for districts statewide and maintain an expansive set of online resources. Nevertheless, there remains room for improvement in providing comprehensive instructions about using DEWS as a tool. This conjecture is plausibly part of the picture. Yet, it could benefit from a deeper study of educators in situ. 

Insights derived from analyzing the effectiveness of the DEWS system point to concrete and actionable improvements in how educational interventions are allocated within Wisconsin public schools. Recall that EWS aim to efficiently target the scarce educational interventions available to schools to students who could benefit from them. Assuming that interventions change outcomes by any value in the range estimated by our causal analysis, we prove that a more straightforward and cheaper targeting mechanism would have yielded at least as significant an improvement in dropout reduction as the DEWS system.

The basis of this claim is a robust statistical pattern regarding the strength of \emph{environmental features} for predicting dropout risk. Environmental or community-based features describe the properties of schools and their districts rather than individual students. For example, environmental features include school size, financial budget, and districts' aggregate socioeconomic and demographic statistics. Individual features, by contrast, include students' test scores, number of school absences, and disciplinary action history. DEWS uses both individual and environmental features to make predictions about individual students.

Given the strength in environmental features, the question remains: Are individual risk scores necessary for effectively targeting interventions? 
Our analysis shows that if we already know these environmental features, incorporating individual features into the predictive model only leads to a slight, marginal improvement in identifying future dropouts. 
Working from weak assumptions, we argue that predicting dropout only using environmental features is sufficient to
identify a cohort of students that are at least as suitable for intervention as the cohort identified by the individualized DEWS scores. That is, intervening on students identified as being at high risk by this alternative, environmental-based targeting strategy would have the same aggregate effect on high school graduation rates in Wisconsin as the individually-focused DEWS predictions.

However, taking a more holistic view of the problem, there are other significant benefits to a community-based allocation strategy. It is simpler to implement and relatively inexpensive: It requires little more than basic information about school districts without collecting fine-grained individual features. Furthermore, it is easier to communicate and justify: There is no need to interpret or explain subtle, hard-to-define concepts such as an ``individual probability of dropout'' that fails to point to a direct and actionable solution. 
Directly targeting low-performing schools needing additional resources is a mechanism that experts and non-experts can readily understand.

\subsection*{Summary of our Method and Analyses}

We first do a deep dive into the prediction model to establish that the risk assessments made by the system accurately sort students by their dropout risk. Over the past decade, nearly 97\%  of students identified as low-risk by DEWS graduated from high school on time, while only 70\%  of students in the high-risk group completed their degrees within four years. As a point of comparison, Wisconsin has a statewide graduation rate close to 90\%. This predictive accuracy is in contrast with work highlighting challenges in predicting life outcomes (such as eviction, job loss, and poor educational outcomes) \citep{harcourt2006against, schumacher2011small, salganik2020measuring}.

Furthermore, contrary to several previous studies on prediction systems in social settings \citep{gandy2010engaging, eubanks2018automating, obermeyer}, 
DEWS predictions are, in fact, more accurate for non-White students than White students. However, these disparities in predictive performance imply that the system could systematically prioritize equally at-risk White students over their non-White counterparts. A similar phenomenon holds if we examine system behavior on students who do and do not have disabilities and those who qualify and do not qualify for free or reduced lunch.

We next investigate the causal effect of DEWS on graduation outcomes via a regression discontinuity design. We estimate that assigning students into higher predicted risk categories improves their chances of graduation by a point estimate of 5\%. However, the 95\% confidence interval for this estimated effect includes zero, even if we optimistically compute confidence intervals via the Bootstrap. 
An immediate question about the effect size is whether and how schools use the system. After its roll-out in 2012, schools quickly adopted DEWS. Tracking visits to the online DEWS portal, we find that around two-thirds of schools regularly log onto the platform. Furthermore, usage is concentrated amongst the more populated school districts with below-average graduation rates and a higher percentage of students from marginalized backgrounds. 
The frequency of system use does not explain the above findings to the extent that we can test this with the available data.  

Motivated by the lessons learned evaluating DEWS, the second part of our work envisions a more straightforward targeting mechanism that is just as effective, if not more effective, than the current system. This alternative program builds on a robust statistical pattern pervasive throughout Wisconsin public schools. Namely, environmental features, defined at the level of schools and districts, contain significant signal about dropout risk across the population. Due to the realities of socioeconomic segregation between districts in Wisconsin, within the same school environment, graduation outcomes are only weakly correlated with available individual student features, including race, gender, and test scores. We support this finding with a comprehensive statistical analysis of the strength of environmental prediction and the marginal value of individual risk score. 

What follows from this insight is a substantive argument about targeting interventions. Although prediction from all features is slightly more accurate than predicting from environmental features alone, we show that this accuracy gap is irrelevant when predictions are used as a basis for intervention. The primary goal of a risk score used for targeting an intervention is to identify a group of students suitable for intervention. As in the case of the DEWS program, we assume that any available interventions apply without a loss of treatment efficacy to any student that is not very likely to graduate. In this case, a risk score used for targeting only needs to identify students that are not too likely to graduate without intervention. We demonstrate that this goal is, in fact, reliably met by an allocation of interventions based on environmental features alone.

\subsection*{Implications}

Our findings have immediate implications for the design of early warning systems and the viability of interventions assigned on the basis of statistical risk scores. First, they challenge the belief that dropout is an unpredictable life outcome rather than a foreseeable event. Owing to the predictive strength of environmental, time-invariant features, future graduation outcomes can be reliably forecast early on in life (i.e., when children enroll in school). It also suggests that we can effectively target additional resources to marginalized students from an earlier age than was previously thought viable in the EWS literature.\footnote{Historically, EWS typically generate predictions no earlier than 5th or 6th grade \citep{balfanz2019early}.}  

Second, our findings challenge the belief that increased prediction accuracy always implies welfare improvements. We do not find evidence that the increase in accuracy from using individual risk scores has improved graduation rates beyond what could have been achieved by school-based targeting. More sophisticated, individually-focused statistical algorithms may be of limited value in settings where structural forces drive outcomes. 

In public education, the overarching goal of an early warning system is to serve as an efficient targeting mechanism. It should answer the question: \emph{``Which students need additional help?''} One of our central findings is that sophisticated predictive models may provide little additional insight into this question beyond simply ranking students according to the aggregate performance of their school district. Evaluating this simple alternative to DEWS takes precedence over questions regarding the effectiveness and availability of educational interventions. After all, as we illustrate in detail later on, the relative merits of the simpler alternative depend only weakly on what educational interventions are available to schools. 

Our findings provide a novel statistical and empirical backbone for robust qualitative insight from education researchers, advocates, and policy-makers. Namely, the bottleneck for resolving low graduation rates is not identifying students at high risk of dropping out but, instead, overcoming structural barriers to accessing well-resourced schools and neighborhoods. In the case of Wisconsin, dropout is disproportionately concentrated in working-class, urban districts where the overwhelming majority of students are Black or Hispanic. In fact, 10\% of all state dropouts come from just five schools. Bringing graduation rates within these five schools to the state average would have an outsize impact on improving the statewide graduation rate. Compared to investing in other community-based social service interventions, funding and implementing sophisticated early warning systems without devoting resources to interventions tackling structural barriers should be carefully evaluated in light of these school-level disparities. 

As a final remark, Wisconsin's socioeconomic realities and social patterns drive these insights regarding the value of community versus individual targeting of educational interventions. They may hold in other similar Midwestern states such as Minnesota or Illinois. However, absent further work, we caution against applying these lessons to places with potentially very different social and geographic compositions, such as New York City.
\section{Background on Early Warning Systems}

Early warning systems emerged as a critical public response to the so-called “dropout crisis” of the early 2000s. During this period, there was widespread, bipartisan recognition of alarmingly high numbers of young adults without high school degrees, especially amongst socioeconomically marginalized populations \citep{senate}. These concerns led Congress to pass the 2001 No Child Left Behind Act, which, amongst its other consequences, increased data collection on students, schools, and districts. 

Following this legislation and increased data availability, EWS boomed in use after the publication of two pilot program studies in 2007 \citep{allensworth2007matters, balfanz2007preventing}.
These studies demonstrated the possibility of surfacing high-fidelity dropout indicators amongst low-income, majority Black and Hispanic public schools in Chicago and Philadelphia. A primary outcome of these studies was the creation of the so-called ``ABC'' indicators of dropout (attendance, behavior, \& coursework), which the authors argue are both simple to measure and can accurately identify students at high risk of dropping out \citep{bruce2011track}. Following these pilot programs, EWS became mainstream: 52\% of the respondents to a survey organized by the U.S Federal Department of Education said they had implemented some version of an EWS by the 2014-15 academic year \citep{survey}.

Over the last decade, there have been various observational studies and randomized control trials evaluating the efficacy of EWS. In 2014, the U.S. Department of Education conducted a randomized control trial with 73 schools containing roughly 38,000 students from three Midwestern states \citep{faria2017getting}. Approximately half of the schools were randomly assigned to implement an EWS. After a year of study, the authors found that schools in the treatment group had a small reduction in chronic absenteeism. However, they observed no effects across other measured outcomes, such as the number of students suspended or the fraction of students with relatively low grade point averages. Notably, the authors point out that schools experienced significant challenges in adopting and implementing these programs.
\cite{mac2019efficacy} conducted a similar randomized control trial across 41 schools over two years, reaching similar conclusions. 

Despite the lack of strong empirical backing, some experts remain optimistic about the efficacy of EWS, and several states have continued to invest in their development (e.g., \href{https://www.doe.mass.edu/ccte/ccr/ewis/}{Massachusetts}). 
Furthermore, these programs have expanded from high schools to  universities \citep{plak2022early} and massive online courses \citep{brooks2015you}.
Summarizing the consensus in the field, \cite{balfanz2019early} state that given the short time horizons and small sample sizes, ``overall, evidence gathering is still in the early stages, promising but not fully confirmed.” Using state-wide data on over 200,000 students from 2013 until 2021, our work addresses this sizable gap in the literature as the first large-scale evaluation of the long-term effect of EWS on the likelihood of on-time high school graduation. 

\subsection{Wisconsin Public Schools and the DEWS Program}

Wisconsin's high schools have one of the highest graduation rates in the United States, with around 90\% of students, on average, receiving their high school diplomas within the expected four years. At the same time, the state also has one of the largest disparities in graduation rates across different demographic groups. For instance, in the 2019-2020 academic year, over 94\% of White students graduated from high school on time, compared to the 75\% graduation rate among Black students.\footnote{See \href{https://nces.ed.gov/ccd/tables/ACGR_RE_and_characteristics_2019-20.asp}{here} for a complete breakdown of graduation rates by state and demographic groups.} There are a total of approximately 65,000 students per grade in Wisconsin public schools, with most of the Black and Hispanic populations concentrated within the larger, urban districts.\footnote{See \href{https://dpi.wi.gov/sites/default/files/imce/eis/pdf/schools_at_a_glance.pdf}{the DPI website} for more information on school demographics.} 

Launched in 2012 by the Department of Public Instruction, the Dropout Early Warning System (DEWS) is part of a broader effort by the state to address these educational disparities and improve graduation rates overall. To do so, DEWS estimates the likelihood that each public middle school student (grades 5 through 8) in Wisconsin will graduate high school on time. DPI generates these scores at the beginning of each academic year and publishes them online via WISEDash, an online platform where school administrators can voluntarily log on to see the output for their students. As discussed previously, website visit statistics provided by DPI show that over two-thirds of districts regularly use the DEWS system. Furthermore, usage is higher amongst lower-income districts with higher percentages of Black and Hispanic students. These districts also tend to have graduate rates below the state's average. For further details on system usage, see the supplementary material (\sectionref{section:usage}).

Predictions are primarily used by school counselors in Wisconsin public high schools as a means of triaging new student cohorts as they enter 9th grade \citetext{\textcolor{DarkGreen}{Personal Communication, Koennitzer, 2022}}. The department recommends---but does not mandate---that schools focus their resources on students assigned to the high-risk category before evaluating students in the lower-risk categories. See the \href{https://dpi.wi.gov/sites/default/files/imce/dews/pdf/DEWS%20Action%20Guide%202015.pdf}{DEWS action guide} for a more comprehensive description of the DPI recommendations regarding how to use and interpret DEWS predictions.

The system generates predictions using over 40 student features. These encompass a wide range of areas, including demographic and socioeconomic information (e.g., race, gender, family income), academic performance (e.g., scores on state standardized exams), as well as community-level statistics  (e.g., percent of cohort that is non-White and school size). The supplementary materials contain a more comprehensive description of the features. The system has two main outputs: 1) the DEWS risk category (or label) and 2) the DEWS score. The DEWS category takes one of three values---low, moderate, or high---representing a discrete assessment of dropout risk. These labels are the main output of the system used to triage students. On the other hand, the DEWS score is an estimated probability of on-time graduation and takes continuous values between 0 and 1.

New models are trained every academic year for every grade. The models are ensembles of state-of-the-art supervised learning methods for tabular data, such as random forests and gradient-boosted decision trees. Models are fit via empirical risk minimization on a dataset of a historical cohort of students. These datasets consist of features and outcomes for the five most recent cohorts of students in a specific grade and for whom graduation outcomes have been observed. For example, the 8th-grade model for the 2020-2021 academic year is trained using data from the 8th-grade cohorts between the 2011-2012 and 2015-16 academic years.

For further background on the DEWS system, see the initial white paper on the program, \citet{knowles2015needles}, or visit the \href{https://dpi.wi.gov/wisedash/districts/about-data/dews}{Wisconsin DPI website}.

\subsection{Evaluation Framework}

From an algorithms point of view, it is helpful to think of an EWS as a sorting procedure. The system takes as input the population of students in the school system and outputs a sorted list (i.e., ranking) of these students according to their dropout risk score. As is common in other EWS, the DEWS scores serve as a concrete proxy for the subjective notion of need for educational interventions. 

Given this ordering, the second key component of an EWS is to choose a threshold. The system prioritizes students whose probability of graduation is below a set threshold to receive interventions over those above this threshold. In the DEWS program, this thresholding is implemented by the DEWS label. As we will explain later, we determine the DEWS categories--low, moderate, and high---by thresholding these continuous DEWS scores. According to the \href{https://dpi.wi.gov/sites/default/files/imce/dews/pdf/DEWS%20Action%20Guide%202015.pdf}{DEWS action guide}, the DPI recommends that schools take immediate action on students who receive a high-risk prediction. Students who receive a moderate risk prediction should only be examined \emph{after} students in the high-risk category. Approximately 11\% of students across the state each year receive a high-risk label, while 6\% receive a moderate-risk label. The remaining students are labeled low risk. 

In light of this sorting analogy, an effective EWS should have two main criteria for success. First, it should accurately rank students according to their intervention needs. That is, students with a lower probability of on-time graduation should be ranked before students with relatively higher probabilities. Such a ranking should also continue to be accurate when restricted to students belonging to historically disadvantaged groups. 
Second, the thresholding and subsequent assignment of students into different risk categories should lead to meaningful changes in individual students' graduation likelihood. Students whose risk scores lie just below the threshold are explicitly prioritized for additional attention by the school over students whose scores lie just above the threshold. High-risk predictions should therefore be self-negating prophecies. Changing the risk label from moderate to high for a given student should \emph{increase} their likelihood of on-time graduation. We will use this evaluation framework to organize our analyses of the DEWS system. 
\section{Does DEWS Accurately Identify Dropout Risk?}
\sectionlabel{section:predictions}

Researchers often discuss early warning systems based on these systems' ability to differentiate between students by their risk of dropping out of high school. As noted, DEWS outputs dropout risk scores from 5th to 8th grade. Our analysis will focus on the performance of the 8th-grade DEWS model, as this is the main output that educators in Wisconsin use to guide their decision-making \citep{personalcommunication}.  


\begin{figure}[t!]
\begin{center}
\includegraphics[width=.48\textwidth]{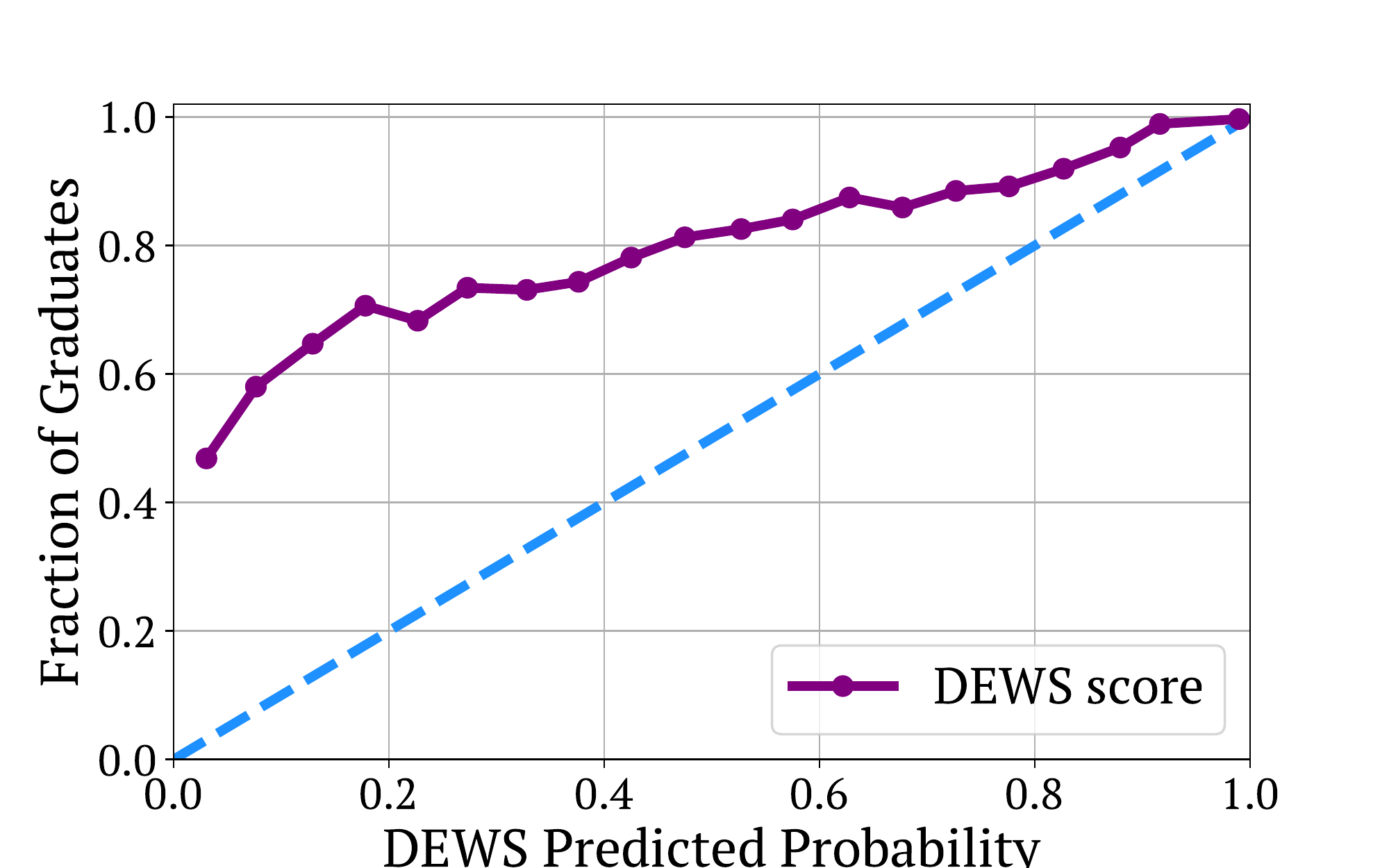}
\includegraphics[width=.48\textwidth]{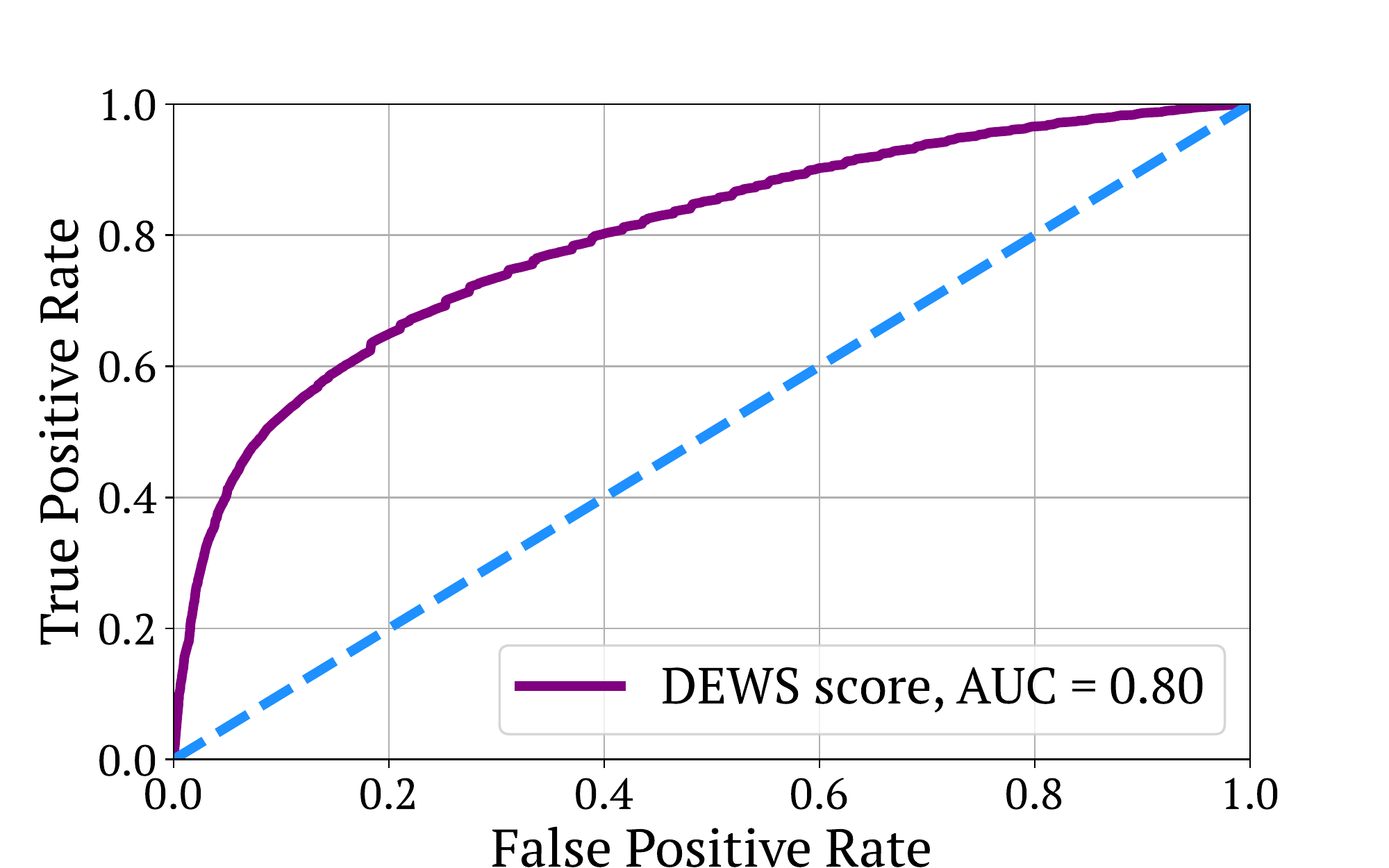}
\end{center}
\caption{\textbf{Left:} calibration curve for the 8th grade DEWS scores (predicted probabilities of on-time graduation) from 2018-2021. \textbf{Right:} ROC curve for the same set of predictions and outcomes. We see that the scores are an accurate assessment of the relative high school dropout risk for students in Wisconsin.}
\figurelabel{figure:overall_accuracy}
\end{figure}


Starting from the 2013-14 academic year--- the first year for which DEWS scores are available---until 2021, there are four cohorts of 8th grade students for whom predictions were generated in 8th grade and we observe graduation outcomes. These cohorts constitute roughly 215,000 students in total.  We focus our discussion on the extent to which DEWS has been predictive for this population. 	

We find significant differences in graduation rates amongst students in different predicted risk categories. Nearly 97\% of the low-risk category students graduate high school on time, while fewer than 70\% of students in the high-risk category do so. This rate is 83\% for students in the moderate-risk category. For context, the overall graduation rate in Wisconsin is around 90\%. These gaps show that DEWS risk labels categorize students into qualitatively meaningful risk groups.			

Examining the continuous DEWS score provides a complementary perspective. In \figureref{figure:overall_accuracy}, we present the calibration curve for these scores. The $x$-axis indicates the DEWS prediction regarding students' likelihood of graduating on time, while the $y$-axis measures the corresponding fraction of students who actually graduate on time. A predictor is calibrated if, amongst the students with a predicted graduation probability of .90, roughly 90\% of them graduate. Graphically, a predictor is perfectly-calibrated if its calibration curve lies on the diagonal $y=x$ line. 

DEWS scores are miscalibrated; they consistently understate the true probability of on-time graduation. However, this miscalibration is benign since the DEWS scores are rank-preserving. Students with higher predicted probabilities have higher graduation rates than those with lower predicted probabilities. That is, the calibration curve is roughly a line with a positive slope. As a result, while the continuous scores are a misleading measure of \emph{absolute} risk, they do provide an adequate assessment of \emph{relative} risk. Viewing DEWS as a sorting algorithm, relative risk is a more relevant metric given its use in this setting. The typical workflow for DEWS is that schools first rank students according to their scores and, in theory, focus their resources on students with the highest risk scores. Accurately measuring relative risk ensures that we appropriately sort students and that those who need more help are more likely to receive attention first. 

We can also look at the induced receiver operating characteristic (ROC curve) to examine the predictiveness of the DEWS scores. This curve describes the complete set of possible true and false positive rates achievable by thresholding the continuous DEWS score. 
The scores achieve a historical AUC (area under the curve) of .8. While lower than the initial AUC estimate of .86 predicted by earlier work on DEWS \citep{knowles2015needles}, these statistics further illustrate how DEWS scores are a non-trivial predictor of dropout. 

\subsection{Disparities in Predictive Accuracy}
\sectionlabel{section:fairness}
\begin{figure}[t!]
\begin{center}
\includegraphics[width=.49\textwidth]{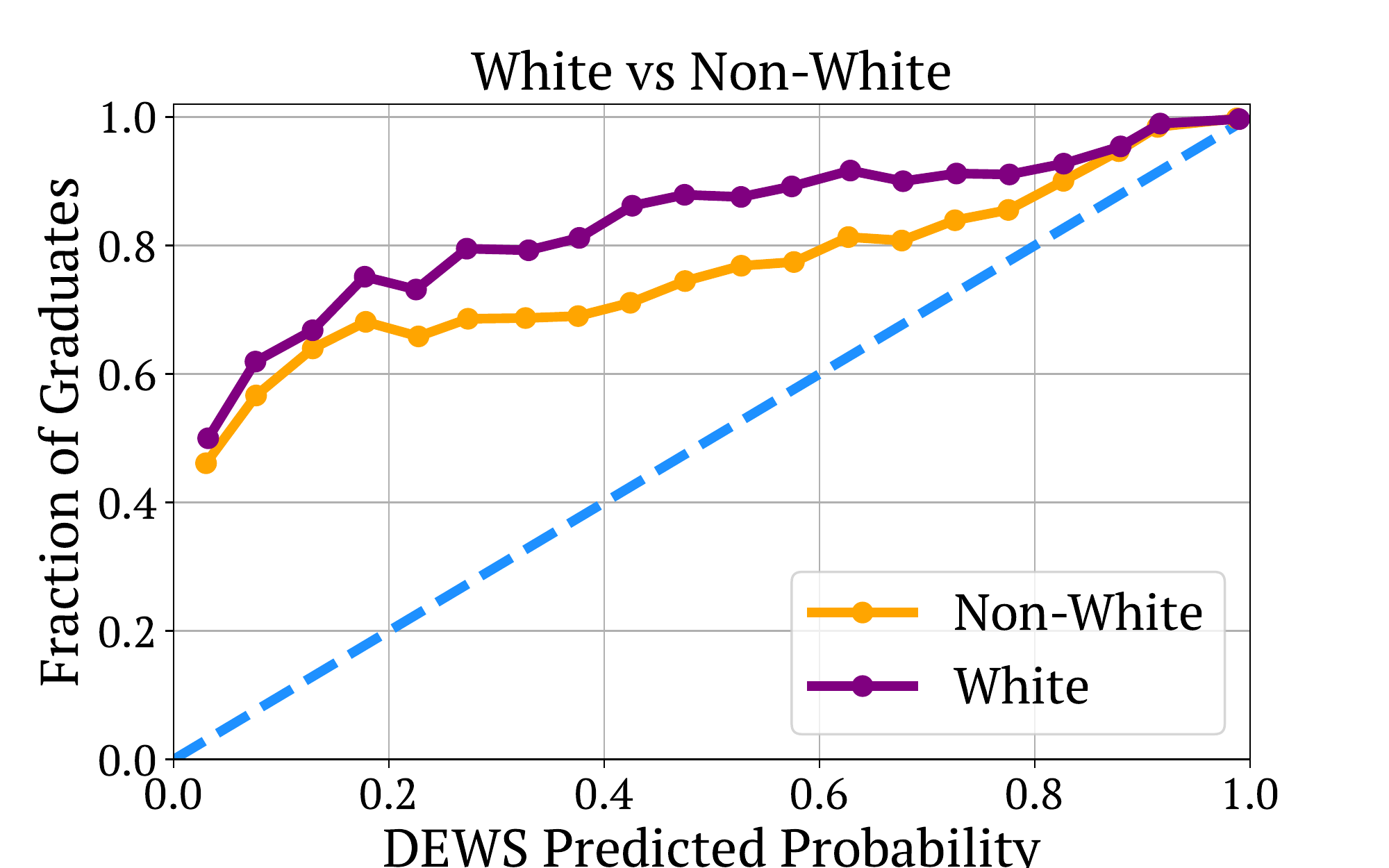}
\includegraphics[width=.49\textwidth]{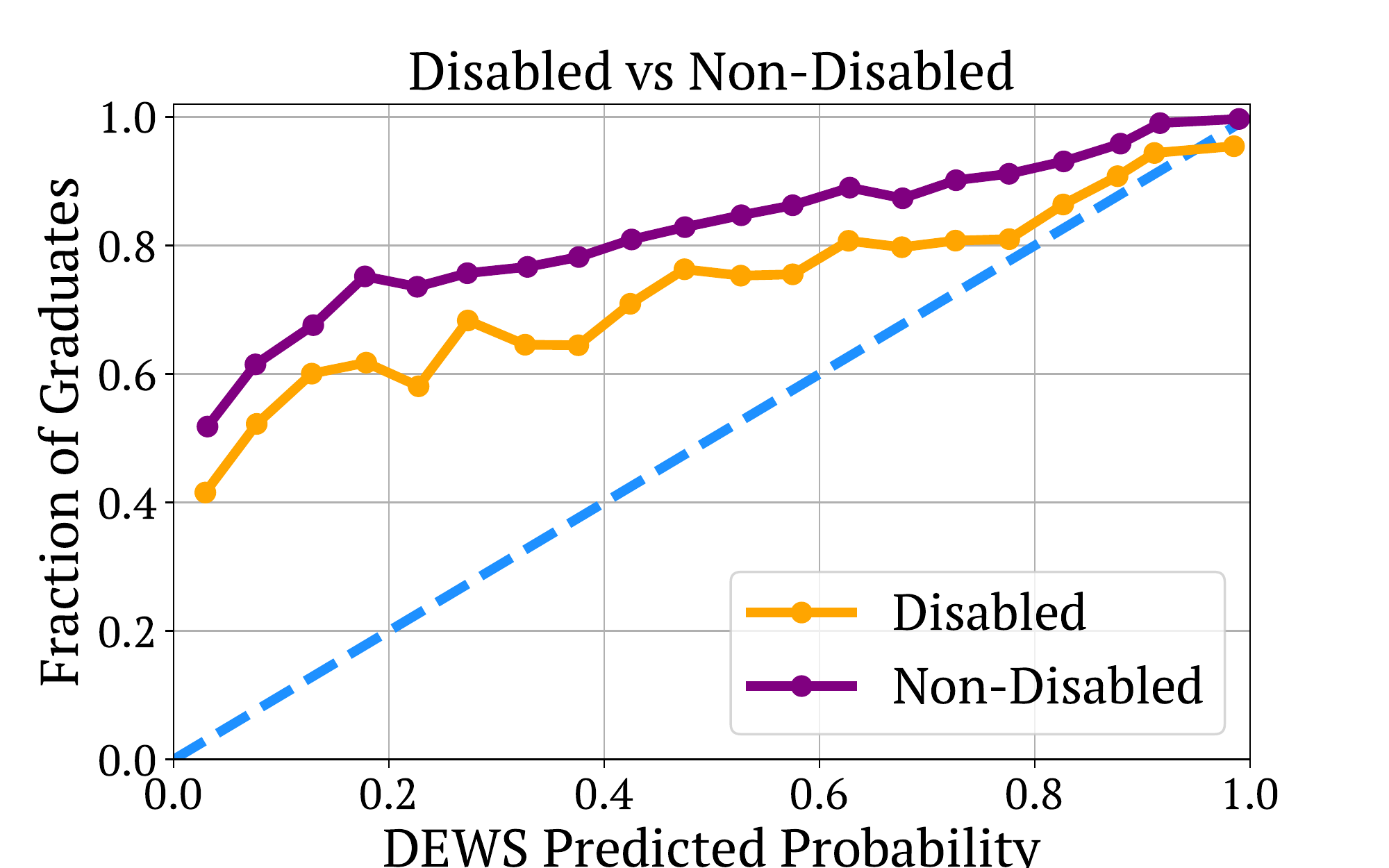}
\includegraphics[width=.49\textwidth]{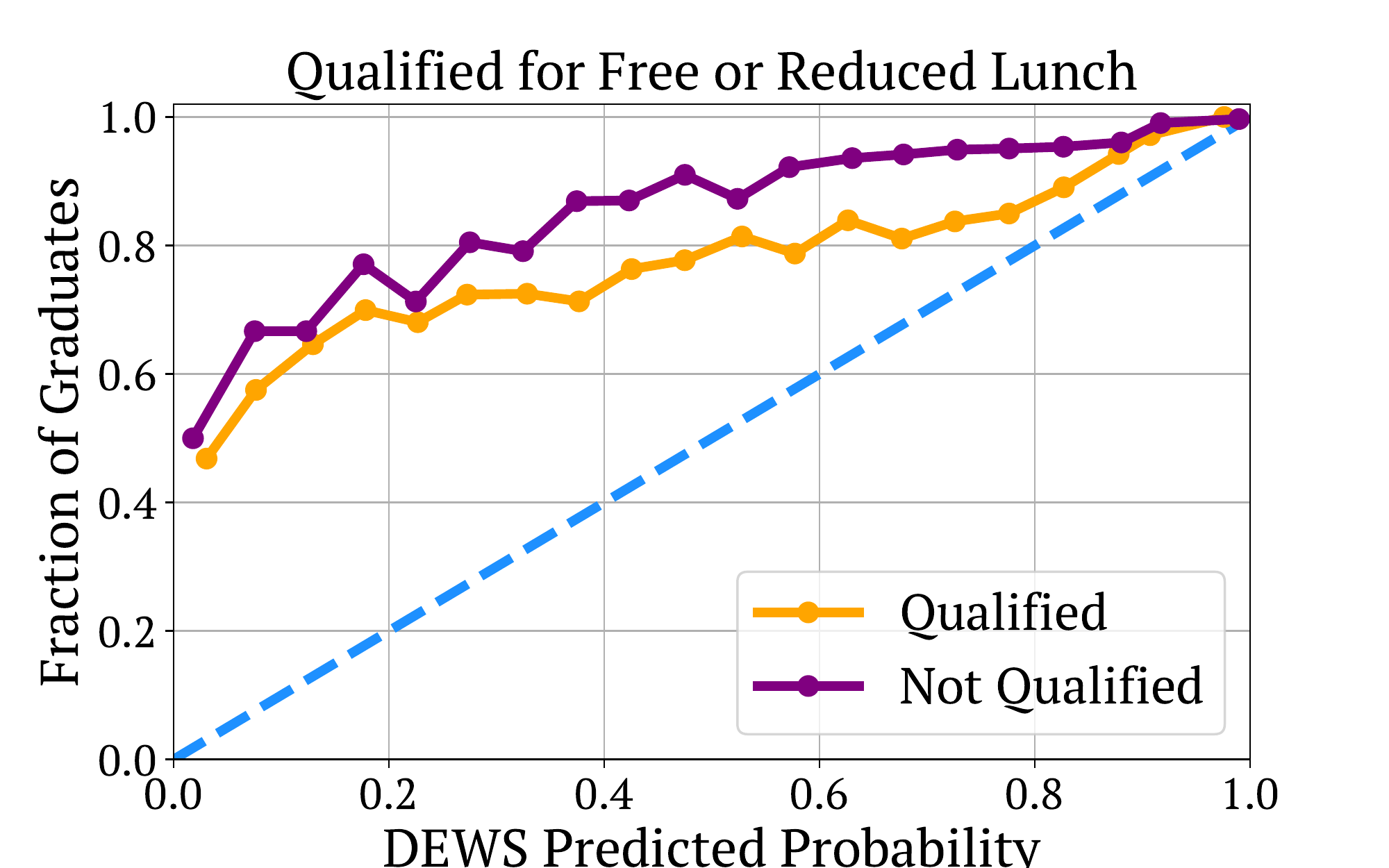}
\includegraphics[width=.49\textwidth]{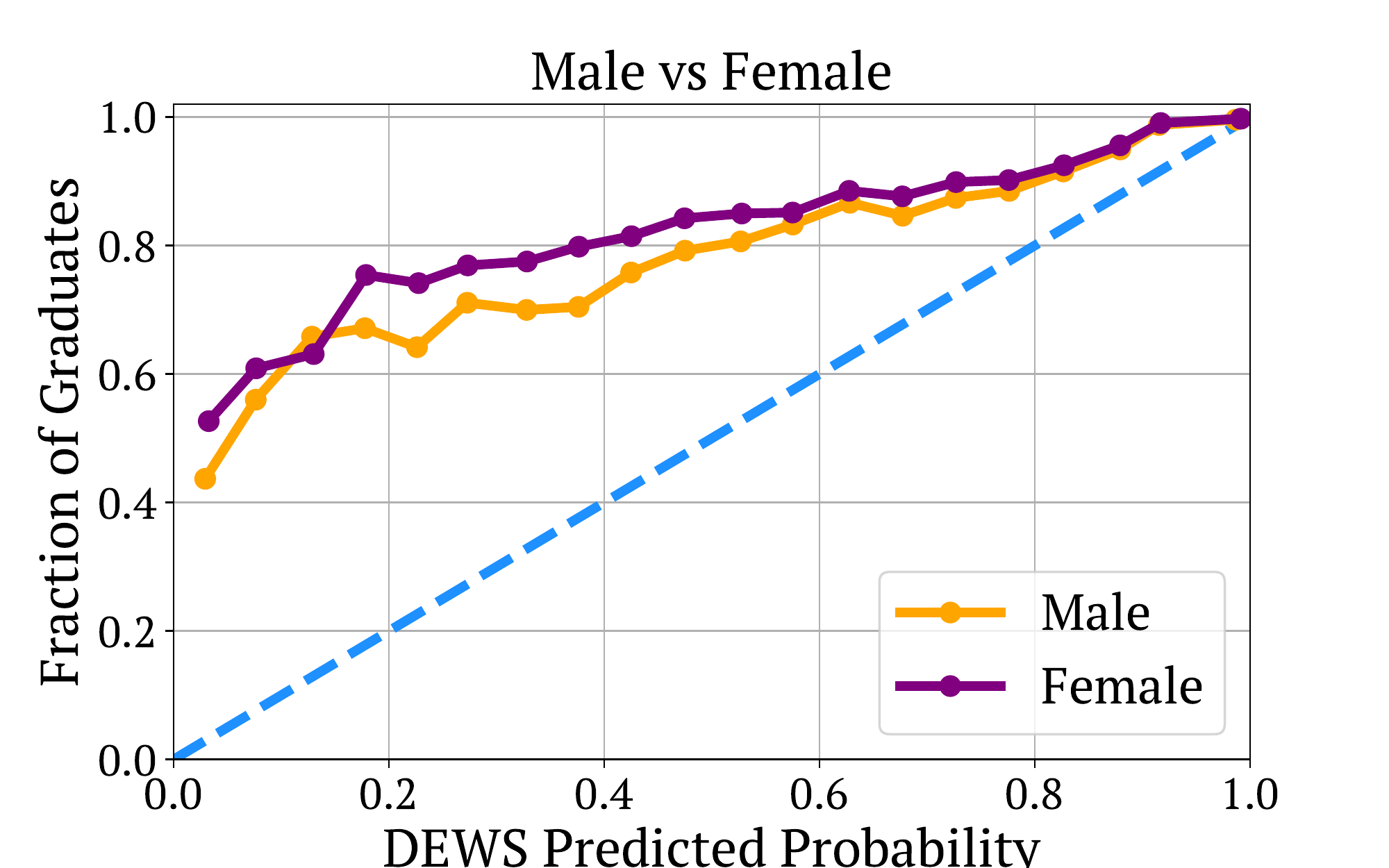}
\end{center}
\caption{Comparing calibration curves for DEWS scores amongst different student groups. \textbf{Top Left:} comparison between White versus non-White students. \textbf{Top Right:} comparison between students diagnosed to have disabilities versus those who have not. \textbf{Bottom Left:} comparison based on students who qualify for free or reduced lunch versus those who do not. \textbf{Bottom Right:} calibration comparison between male and female students.}
\figurelabel{figure:disparities}
\end{figure}

Our results thus far establish that DEWS meaningfully sorts students students by their dropout risk, when evaluated across the entire student population. However, one of the stated goals of the system is to improve graduation outcomes specifically for students from marginalized groups \citep{knowles2015needles}. We are therefore also compelled to evaluate DEWS by restricting our analysis to students from such groups. 

Our investigation into these concerns reveals mixed results. On the one hand, from the plots in \figureref{figure:disparities}, we see that predictions for various historically marginalized groups have lower calibration errors: The calibration curve for non-White, low-income students, and students with disabilities is relatively closer to the $y=x$ diagonal. Hence, DEWS scores provide a \emph{more} accurate assessment of these students' absolute dropout risk. On the other hand, from the perspective of relative risk, if schools select students for intervention by ranking their scores and selecting those with the lowest predicted probability of graduation, students from historically disadvantaged groups would be systematically de-prioritized for additional school attention.

To see this, consider the top left plot comparing predictions on White versus non-White students.\footnote{The precise comparison here is between the group of students in the dataset with a 1 in the column indicating whether a student is White and the group for which this feature is equal to 0. As per DPI practices, if this feature is 0, a student could be Black, Hispanic, Asian, Pacific Islander, Native American, or belong to two or more races. We use non-White to refer to students with corresponding 0 values in the "is White" column. } 
Amongst the population of students whose true rate of on-time graduation was~.8 (i.e., the $y$-axis value equals~.8), White students were assigned a DEWS score of around~.35 while non-White students had an assigned DEWS score closer to~.6. Recall that DEWS scores predict the probability of on-time graduation: Lower scores indicate higher predicted risk. While both groups of students are at equal risk of dropping out, given that they share the same value on the $y$-axis, if schools assign interventions based on ranked DEWS scores, White students would systematically receive interventions before non-White students. Furthermore, because the calibration curves for non-White students lie consistently below the curves for White students, this observation holds across the entire spectrum, not just for this 80\% group.

It remains unclear if DEWS has exacerbated existing inequalities in graduation outcomes between the demographic groups we consider in \figureref{figure:disparities}. Nevertheless, these disparities in accuracy are worth noting. Institutions building these predictive systems should fix these disparities when training their predictors. Recent advances from the algorithmic fairness literature can help address the miscalibration issues we observe \citep{hebert2018multicalibration, pfisterer2021mcboost}.

\section{Does DEWS Improve Graduation Outcomes?}
\sectionlabel{sec:interventions}

The previous section establishes that DEWS scores appropriately rank students by relative risk. Given the explicitly stated goal of DEWS to improve graduation rates, it remains to show whether thresholding these scores into discrete risk categories translates into the desired outcome. 

Receiving a high-risk label ought to be a self-negating prophecy. Students in this group get priority in receiving educational interventions compared to the counterfactual setting, where they received a moderate or low-risk label. Increasing a student's predicted level of risk should \emph{increase} the likelihood that they graduate from high school on time. 

To be precise, the exact intervention whose effect we aim to measure is assigning students to a higher-risk bucket. According to \href{https://dpi.wi.gov/sites/default/files/imce/dews/pdf/DEWS%20Action%20Guide%202015.pdf}{DEWS action guide} signaling that a student is at high risk may trigger a host of educational interventions (e.g., counseling, tutoring, and so on) that may differ between schools across the state. Hence, while the intervention is well-defined, it may have heterogeneous effects on different groups of students. We explicitly test for this possibility later on.

On a technical level, we measure the impact of assigning students to higher-risk categories via a regression discontinuity design (RDD).
These are quasi-experimental methods commonly used to estimate treatment effects within the econometrics and causal inference literature \citep{angrist1999using,imbens2008regression,thistlethwaite1960regression}.
The key insight enabling this approach is that DEWS generates the discrete risk categories by thresholding the DEWS score, a smoothly-varying and continuous variable. Intuitively, this discontinuity serves as a ``natural experiment'' amongst the subset of students whose relevant statistics lie close to the threshold. We refer the reader to \cite{imbens2008regression} for a more formal background on RDDs.

In more detail, as part of its prediction pipeline, given student features $x$, DEWS generates an error estimate $e(x)$ around the DEWS scores $p(x) \in [0,1]$. This error estimate is used to compute adjusted scores $(\ell(x), u(x)) = (p(x) - e(x), p(x) + e(x))$.\footnote{Adjusted scores are clipped to the unit interval.}  The error estimates $e(x)$ can, in principle, vary with x. However, due to the system's idiosyncrasies, they are nearly constant and equal to $\sim.03$ for the vast majority of students (see \figureref{figure:bin_width_histogram}). Therefore, the reader can think of the adjusted scores to be the DEWS score shifted by a constant, $(\ell(x), u(x)) = (p(x) - .03, p(x) + .03)$.

DEWS risk categories are determined as follows. If the upper score $u(x)$ is below a department-chosen threshold of $\ts=.785$, then students are assigned a high-risk label.\footnote{The DPI chose the $\ts=.785$ threshold early on in the DEWS program to appropriately balance the number of students assigned to each risk category.}
On the other hand, if the lower score is above $\ts$, then students are assigned a low-risk label. Lastly, students are assigned to the moderate-risk category if $\ts$ is contained in the interval $(\ell(x), u(x))$. 
The intuition behind this thresholding process is that upper adjusted score is low, then students are predicted to be unlikely to graduate on time and should hence be prioritized into the high-risk category. The opposite rationale holds for the low-risk category. 

RDDs are motivated by the observation that risk categories (i.e., the treatment) for students whose predicted confidence bounds are close to the threshold are essentially random. 
By looking at students whose upper score $u(x)$ is just around this threshold $\ts$, we can infer the expected difference in the likelihood of on-time graduation that results from changing a student’s DEWS label from moderate to high-risk. Similarly, comparing students whose lower score $\ell(x)$ is around $\ts$ reveals the analogous impact of assigning students to the moderate versus low-risk category. 

We use notation from the potential outcomes framework to introduce our analysis more formally. We let $Y(r) \in  \{0,1\} $ be the indicator variable denoting on-time student graduation under assignment into predicted risk category $r$, where $r$ can take values in the set \{low, moderate, high\}, and let $Y$ denote the observed historical outcome. We use $x$ to denote the vector of student features and  
$(\ell(x), u(x)) \in [0,1] \times [0,1]$ to denote the lower (respectively, upper) adjusted scores determined by the DEWS system for each student. We assume that the conditional expectations of outcomes are smooth, continuous functions of these confidence bounds.


\begin{figure}[t!]
\begin{center}
\includegraphics[width=.48\textwidth]{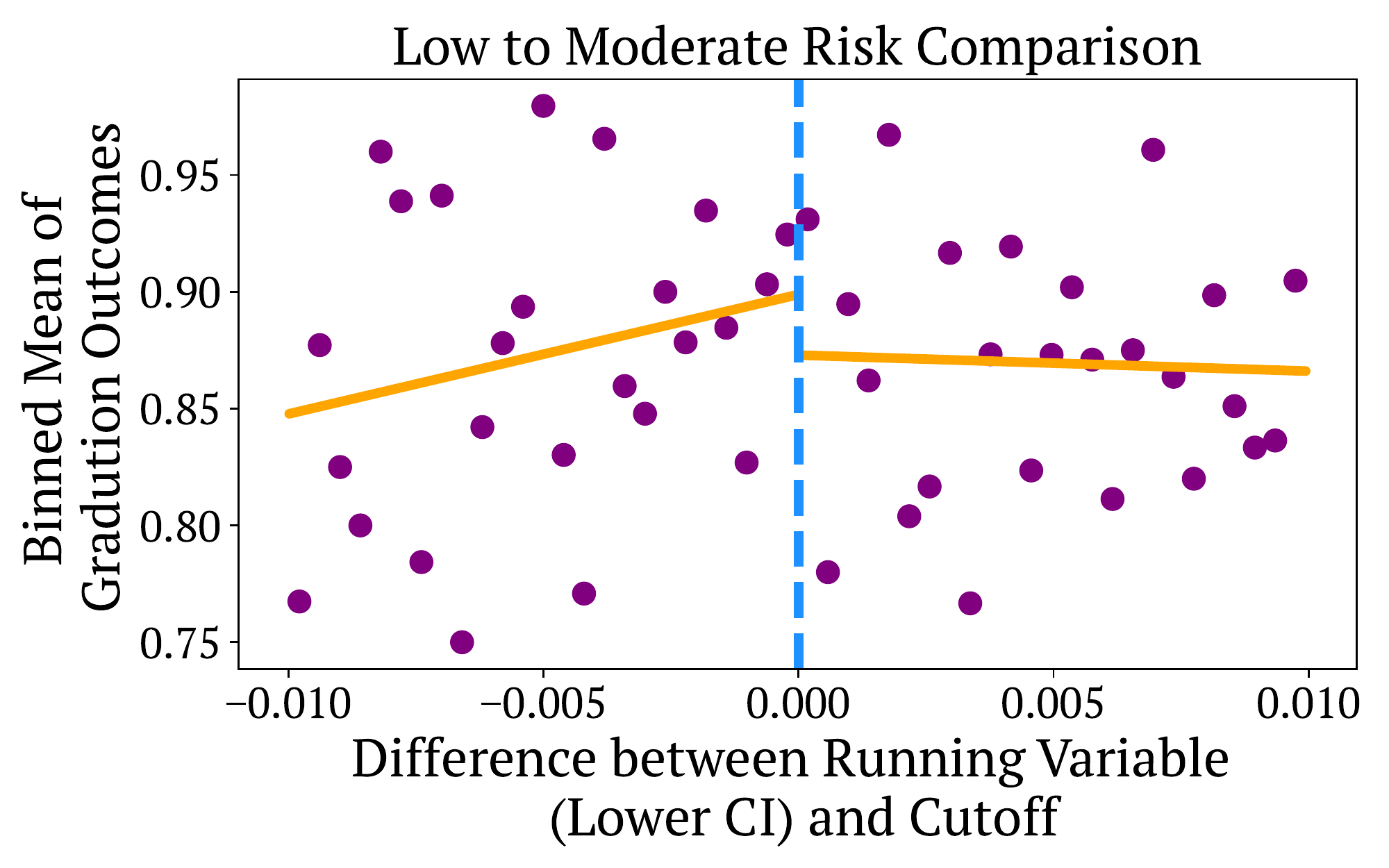}
\includegraphics[width=.48\textwidth]{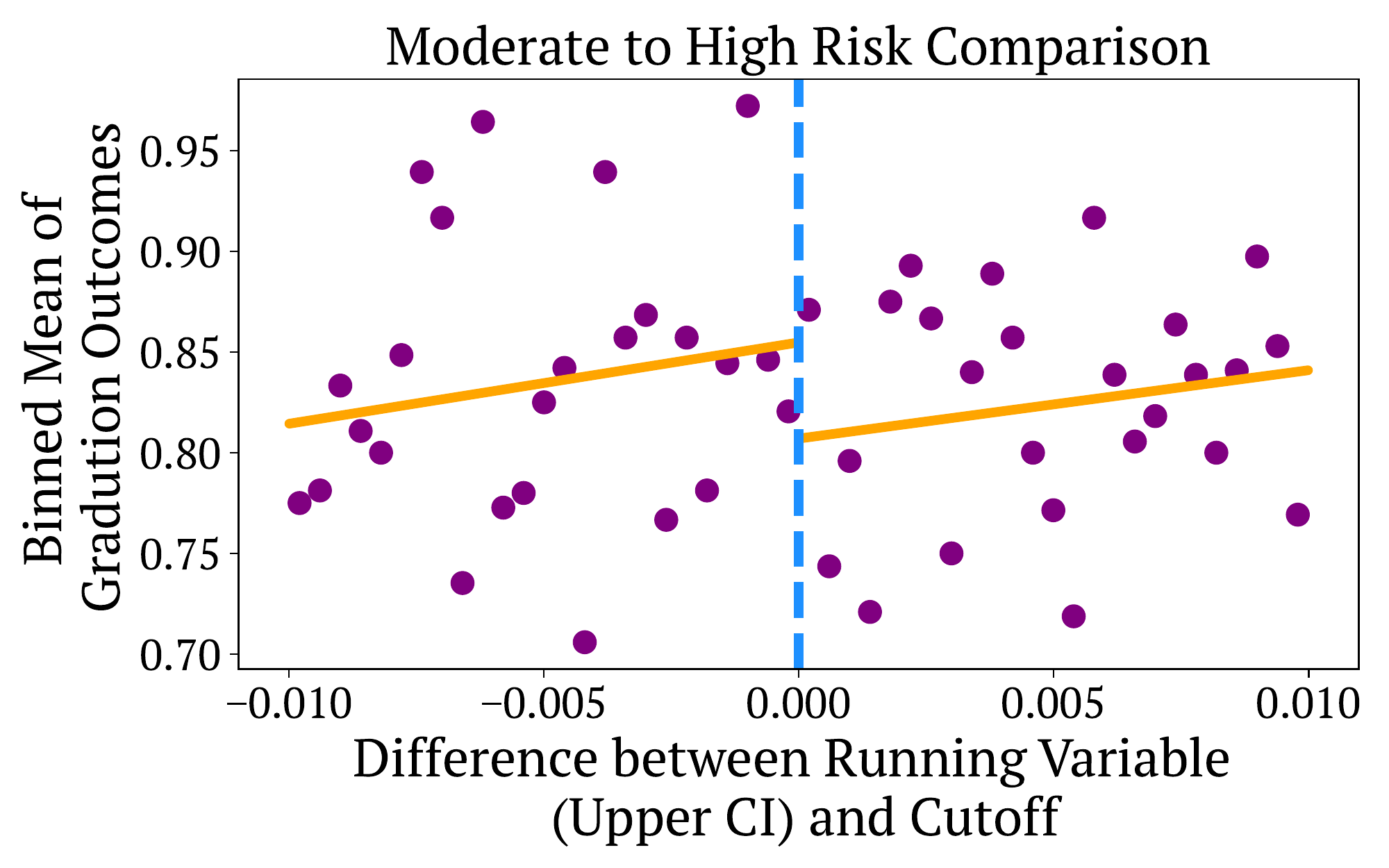}
\end{center}
\caption{ Visualization of RDD results. Purple dots correspond to the average graduation outcomes for students within each confidence interval bin. Yellow lines indicate fitted values from performing local linear regression. The point estimate for the treatment effect is the gap between yellow lines at the point when they intersect the threshold (blue dotted line). \textbf{Left:} Treatment effect of increasing risk category from low to moderate. Students to the left side of the threshold receive a moderate-risk label. \textbf{Right:} Treatment effect of increasing risk category from moderate to high risk. Students to the left side of the threshold receive a high-risk label.}
\figurelabel{figure:rdd}
\end{figure}


\begin{assumption}(smoothness of conditional expectations)
For all values of $r$ in the set \{low, moderate, high\}, the functions $\E[Y(r) \mid \ell(x) = c]$ and $\E[Y(r) \mid u(x) = c]$ are twice continuously differentiable and smooth functions of $c$, for all values $c$ in an open set containing $\ts$.
\end{assumption}

Under this standard technical condition, it is well-known that performing local linear regression around the cutoff value estimates the desired causal effects up to a small bias term. In particular, consider the treatment effect $\tmh$ of increasing the DEWS label from moderate to high risk for students whose predicted upper confidence bound is equal to the threshold:
\begin{align*}
	\tmh  \defeq \E[Y(\text{high}) - Y(\text{moderate})\mid u(x) = \ts].
\end{align*}
This treatment effect is equal to the value~$\btmh$ solving the ordinary least squares objective
\begin{align*}
 \E 1\{|u(x)- \ts| \leq h\} \left(Y - \alpha - \btmh 1\{u(x) < \ts \} - \beta (u(x) - t^*)   1\{u(x) < \ts \}   - \gamma (u(x) - t^*) \right)^2,
\end{align*}
up to bias that is order $h^2$ and statistical error that is order $n^{-1/2}$
\citep{imbens2008regression}. Here, the bandwidth parameter~$h > 0$ ensures that only points close to the cutoff enter the regression, and $n$ is the number of points within that bandwidth. We can solve for the parameters $\alpha, \beta, \gamma,$ and $~\btmh$ via OLS. 

Likewise, if we let $\tml$ be the average difference in graduation rates resulting from increasing the predicted risk category from low to moderate,
\begin{align*}
	\tml  \defeq \E[Y(\text{moderate}) - Y(\text{low})\mid \ell(x) = \ts],
\end{align*}
then this treatment effect is also $O(h^2 + n^{-1/2})$ close to the value $\btml$ that solves
\begin{align*}
	\E 1\{|\ell(x)- \ts| \leq h\} \left(Y - \alpha - \btml 1\{\ell(x) < \ts \} -  \beta (\ell(x) - \ts) 1\{\ell(x) < \ts\}  - \gamma (\ell(x) - \ts) \right)^2 .
\end{align*}

\begin{table}[b!]
\begin{center}
\resizebox{\textwidth}{!}{\begin{tabular}{ccccc}
 Causal Effect &  Point Estimate & 95\% Confidence Interval &  $p$-value  & $n$ \\
\hline
\centering
    $\tml$:\; increasing risk from low to moderate & 0.022 & (-0.027	0.071) & 0.38 & 2746 \\
  $\tmh$:\;  increasing risk from moderate to high & 0.048 & (-0.02,	0.116) & .17 & 1888 
\end{tabular}}
\end{center}
\caption{Causal effect estimates from the RDD. We compute the 95\% confidence intervals using a Normal approximation for the regression coefficients. Computing these confidence intervals via the Bootstrap yields essentially identical values. We present these in \tableref{table:ci_rdd}. The $n$ values correspond to the number of points within the threshold's chosen bandwidth. While the point estimates are non-zero, they come with nontrivial uncertainty. We cannot, with high confidence, rule out the possibility that there is no effect.}
\tablelabel{table:rdd}
\end{table}
\newpage
\begin{figure}
    \centering
    \includegraphics[width=.47\textwidth]{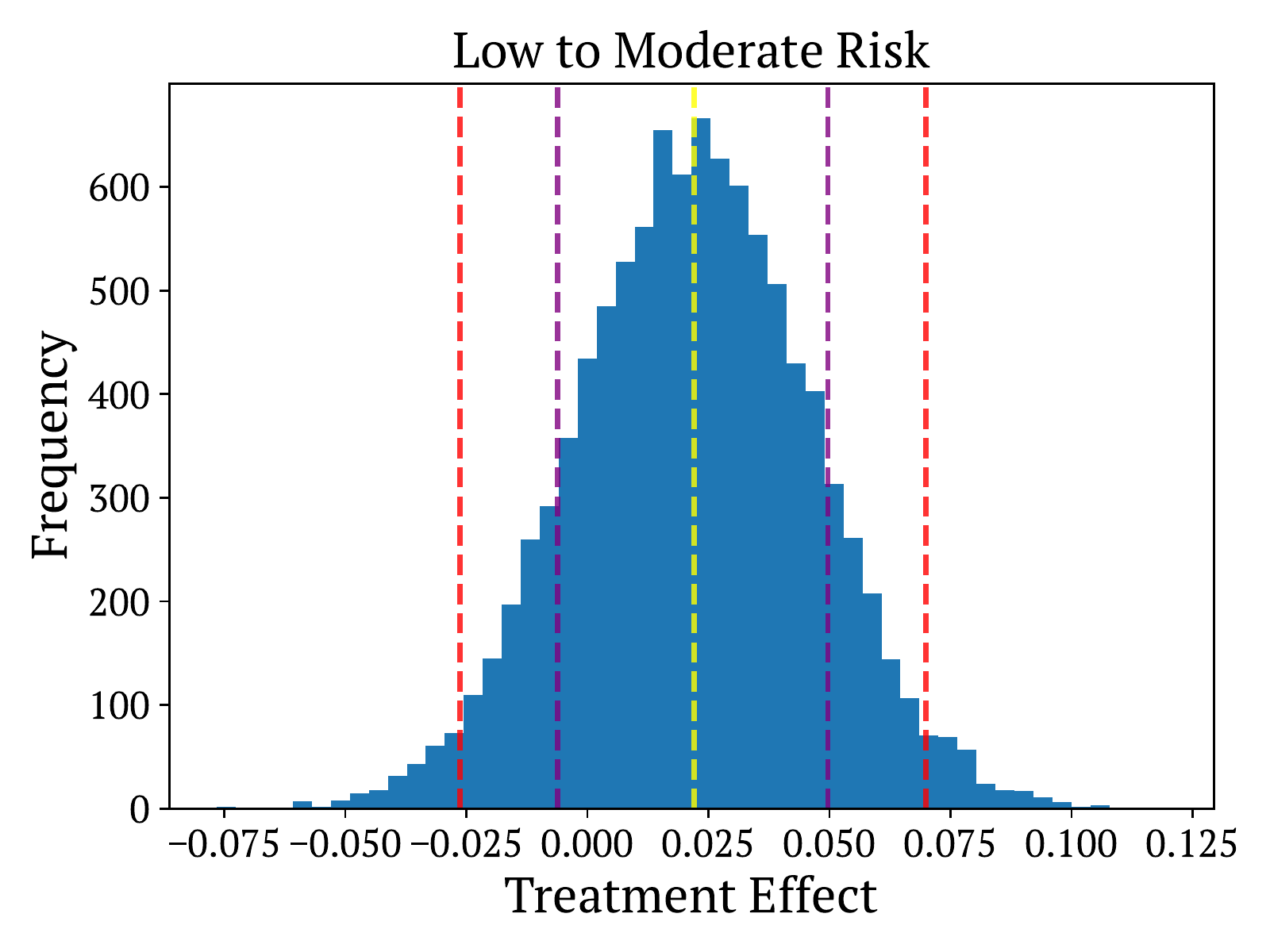}
    \includegraphics[width=.47\textwidth]{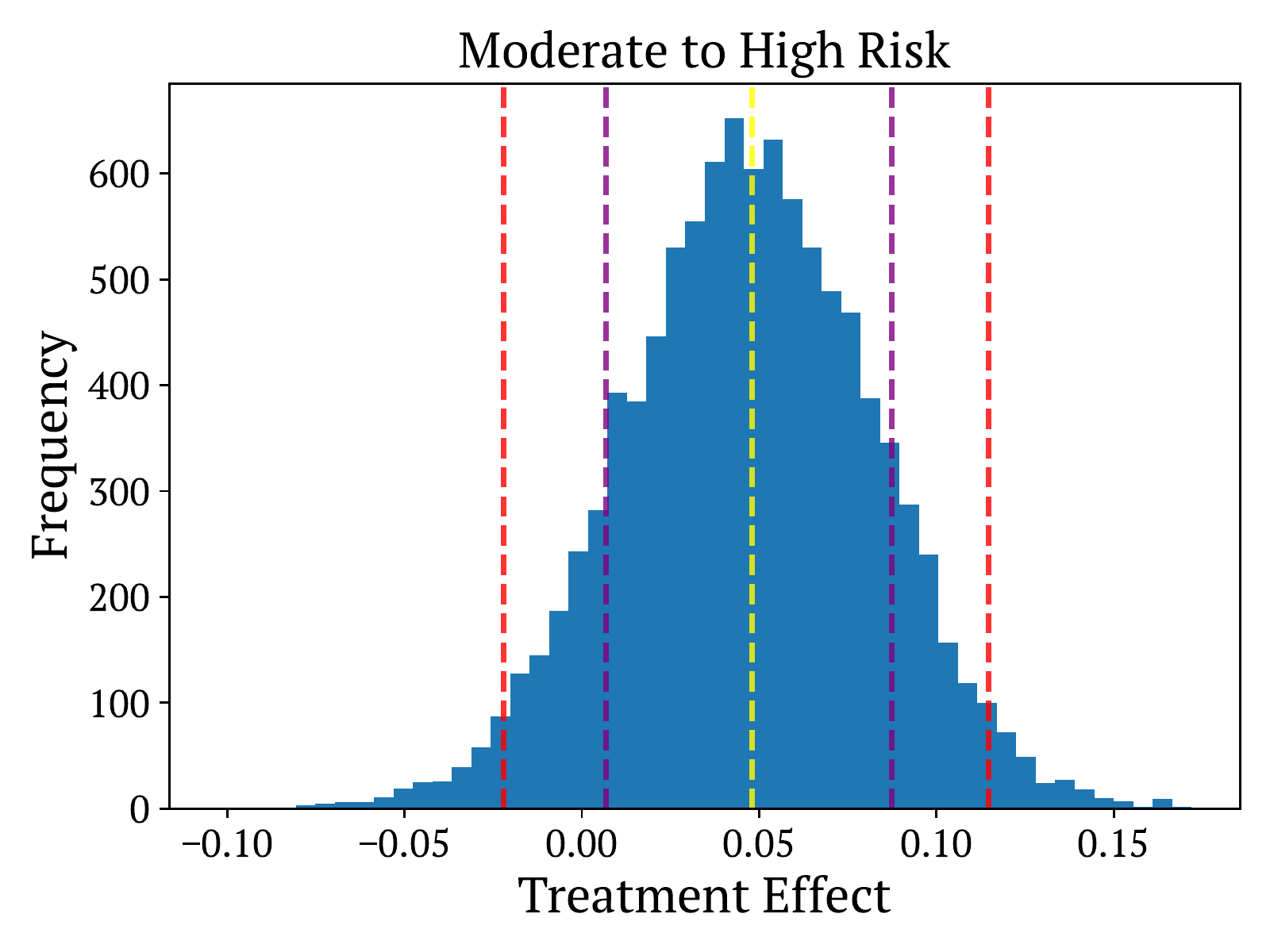}
    \caption{Visualization of statistical uncertainty in the causal effect estimates from the RDD presented in \figureref{figure:rdd} and \tableref{table:rdd}. \textbf{Left:} Histogram of 10,000 Bootstrap estimates of the $\tmh$, the causal effect of increasing the DEWS category from moderate to high. \textbf{Right:} Analogous histogram for $\tml$, the causal effect of increasing risk category from low to moderate. The point estimates for each estimand are denoted in yellow. The 95 and 75 confidence intervals are denoted in red and purple, respectively. See \tableref{table:ci_rdd} for the exact numbers.}
    \figurelabel{fig:bootstrap}.
\end{figure}

\vspace{-5pt}
\begin{table}
\begin{center}
\begin{tabular}{c|c|c}
& Moderate to High & Low to Moderate \\ 
\hline
Point Estimate & 0.048 & .022 \\ 
95\% CI: Bootstrap & (-0.022, 0.115) & (-0.026 0.07) \\
95\% CI: Normal Approx & (-0.02,	0.116) & (-0.027	0.071)   \\
75\% CI: Bootstrap & (0.007, 0.087) & (-0.006, 0.05)\\
75\% CI: Normal Approx & (0.008,	0.0877) &  (-0.007,	0.05)
\end{tabular}
\end{center}
\caption{Numerical values for the various estimates presented visually in \figureref{fig:bootstrap}. Bootstrap confidence intervals are computer by resampling the dataset used to compute the RDD estimates from \figureref{figure:rdd} 10,000 times, using OLS to compute the regression coefficients, and then examining the relevant percentiles. Normal confidence intervals are computed via the standard Normal error approximation to OLS regression coefficients. As is the case for all our analyses, both methods return nearly identical answers.}
\tablelabel{table:ci_rdd}
\end{table}
\clearpage

We present the results of this regression discontinuity analysis in \figureref{figure:rdd} and \tableref{table:rdd}. 
We estimate that changing the DEWS category from moderate to high-risk increases the likelihood of on-time graduation by approximately 5\%. The analogous treatment effect of assigning increasing the predicted risk category from low to moderate risk leads to a smaller increase in the likelihood of graduation of around 3\%. 

For context, this 5\% point estimate is approximately equal to the difference in graduation rates between schools at the 35th and 85th percentiles (of a fraction of on-time graduates) in Wisconsin. If we believe that the likelihood of on-time graduation is almost exclusively determined by the school a student attends, the treatment effect of assigning a student into the high-risk category is approximately equal to the causal effect of transferring a student from a relatively under-performing school (i.e., a school with a graduation rate in the 35th percentile) to an above average school in the 85th percentile. 

However, both of these estimates come with nontrivial uncertainty. The 95\% confidence interval for $\tmh$ ranges from -2\% to 11.6\%. This interval contains 0, and we cannot, at this level of confidence, rule out the possibility that assigning students into different risk categories does not change their likelihood of on-time graduation. The corresponding 95\% confidence interval for $\tml$ reflects the smaller point estimate: It ranges from -2.7\% to 7.1\%. Please see \figureref{fig:bootstrap} for a visual illustration of this uncertainty as indicated by the histogram of Bootstrap estimates.

If we are willing to reduce our confidence level, the 75\% confidence interval for $\tmh$ is 0.8\% to 8.7\% and excludes 0. While both estimates are directionally correct---increasing the predicted risk reduces the probability of dropout---these are not rigorously conclusive and call for further investigation into the impacts of specific educational interventions. We return to this point in the discussion section.
 
We restrict our regression analyses to the population of students whose school districts log on to the portal at least once per year during the period for which we have usage data and choose the bandwidth parameter $h$ to be .01. Our conclusions are robust to the bandwidth parameter. They also hold when restricting our analysis to specific subpopulations, such as non-White students and students who qualify for free or reduced lunch. We present a more comprehensive set of sensitivity and robustness checks of the regression discontinuity design in \sectionref{section:robustness}. 

\paragraph{Further Analyses.} Our RDD aims to infer whether the choice of risk category changes the likelihood of on-time graduation for a particular subset of students, i.e., those who lie close to the thresholds which determine the risk categories. We consider an additional statistical test to identify if DEWS has had a non-zero treatment effect in ways not captured by the DEWS category. The main idea is to see if including DEWS scores as a feature on top of all other features improves accuracy, as a means to detect broader statistical dependencies between DEWS predictions and student outcomes. This identification strategy of ``predicting from predictions'' draws on recent work by \cite{mendler2022predicting}. The test, presented in \sectionref{section:independence_tests}, again shows we cannot with high certainty rule out the possibility that DEWS predictions are ineffective.
\section{A School-Focused Targeting Mechanism}
\sectionlabel{section:environment}

Our causal analysis of DEWS shows that it is plausible, though not exactly certain, that labeling students high risk triggers educational interventions that increase their likelihood of on-time graduation by a single digit percentage. 
In this section, we ask: Assuming that these interventions are indeed effective, is DEWS the best way to target them to students? We argue that, for the entire range of plausible treatment effects, a simpler school-based targeting mechanism would likely have a similar bottom-line impact on the state's graduation rate. 

For the sake of the argument, suppose that there is an intervention that schools can administer to a student that will improve their graduation likelihood by 5\%. We choose 5\% as it is close to the point estimate of our regression discontinuity analysis. Nevertheless, the argument holds for any value in the 95\% confidence interval estimated previously (i.e., -2\% to 12\%).\footnote{On a technical note, the RDD estimates an average treatment effect meaning that the intervention may have different effects on different individuals or groups. However, to the extent that we can test for this in the data, we see no evidence of significantly heterogeneous effects. Restricting the RDD to specific subgroups of students consistently yields results within this -2\% to 12\% range. Please see \sectionref{section:robustness} for further details.}
Administering an intervention to a student is \emph{efficient}, if that student's true likelihood of graduation absent an intervention is well below 95\%. Intervening on students exceedingly likely to graduate results in a diminished treatment effect and wastes scarce resources. Stated otherwise, if $\Pr[i \text{ Graduates}  \mid \text{ no intervention}] > .95$, then $\Pr[ i \text{ Graduates} \mid \text{do(intervention)}]  - \Pr[ i \text{ Graduates} \mid \text{ no intervention}] < .05$, since probabilities cannot go past 1.
In short, a sufficient condition for a risk score to enable efficient targeting of interventions is that it reliably identifies a group of students with  graduation likelihoods well below 95\%.\footnote{This assumption---that a student's suitability for intervention is determined by their likelihood of graduation sans intervention---is core to DEWS and all other EWS that base decisions primarily on students' predicted probabilities of on-time graduation. Moreover, as discussed previously, it matches existing school practices in Wisconsin \citep{personalcommunication}.}

School-based targeting of interventions is at least as good as DEWS because a student's academic environment is a sufficiently strong predictor of their graduation outcomes.
More specifically, a predictor that only uses features defined at the level of schools and districts to forecast student outcomes-- an environmental predictor --is nearly as accurate as DEWS style models that draw upon richer sources of data. Since an individualized risk model like uses \emph{both} environmental and individual level features, it \emph{is} slightly more accurate than an environmental model that does not have access to individual-level covariates (e.g., race or test scores). 
However, this gap is likely irrelevant when prediction is used to target interventions.

Recall that DEWS explicitly prioritizes 11\% of students for intervention (11\% of students are labeled high risk). Since the goal of targeting is to improve graduation rates, the main way of evaluating a targeting strategy is to consider the aggregate impact on the statewide graduation rate that arises from intervening on students with the lowest predicted scores, i.e., those in the bottom 11\%. 
The main technical point of this section is that because the students identified as high risk by the individual predictor, and those identified as high risk by the environmental predictor, both have likelihoods of graduation below 70\%, applying an intervention that increases on-time graduation by an additive 5\% would have the same impact on the total number of students who graduate from high school. Following a brief overview of environmental versus individual predictors, we will expand upon the empirical insights supporting this finding.

\paragraph{Defining Environmental Predictors} Risk models in Early Warning Systems predict a binary indicator of on-time graduation, $y$, given a $d$-dimensional feature vector $x$. Conceptually, we take these $d$ features and partition them into 2 sets: environmental and individual, $x = (x_{\mathrm{env}}, x_{\mathrm{ind}})$. We define a feature as an \emph{individual feature} if it measures information that directly corresponds to a specific student. Examples of these features are variables like race, gender, or the number of days a student attended school. On the other hand,  \emph{environmental features} are those that describe a student’s community. These are variables like the size of a student’s cohort, the
average math score in their school, or the median income in their district.\footnote{Please see the supplementary material for a complete list of features and their relevant categorizations.}
An environmental predictor $f_\mathrm{env}$ predicts the outcome $y$ using \emph{only} the environmental features, $f_\mathrm{env}(x_{\mathrm{env}}) \approx \Pr[y=1 \mid x_{\mathrm{env}}]$. On the other hand, an individualized predictor $f_\mathrm{ind}$ uses \emph{all} the features: $f_\mathrm{ind}(x_{\mathrm{env}}, x_{\mathrm{ind}}) \approx \Pr[y=1 \mid x_{\mathrm{env}}, x_{\mathrm{ind}}]$.

\paragraph{Experimental Setup.} For our analysis, we take the entire dataset of 8th graders for which we observe graduation outcomes and assign 80\% of them ($\approx$160k) to a training set and 20\% ($\approx$40k) to a test set. We partition the available features into two groups: environmental and individual.

On the training set, we train two separate models: one model that forecasts graduation outcomes just using the environmental features (the environmental predictor) and a different model that predicts on-time graduation using \emph{both} individual features and environmental features (the individual risk model). Both training procedures are identical. They are both state-of-the-art learning algorithms for tabular data (gradient-boosted decision trees from the catboost library \citep{catboost}), separately trained on the same  data. They only differ in the subset of features to which they have access to as outlined in the previous paragraph.\footnote{For these experiments, in addition to the features present within the DEWS system, we incorporate easily identifiable environmental features. 
In particular, we include community-level statistics regarding
students’ public school districts drawn from the US Census Bureau’s American Community Survey and the National Center for Education Statistics. These span areas like school expenditures per student and racial demographics in the district.} 

\paragraph{Results.} We evaluate the predictions of both models on the held-out test set of 40k students and illustrate the results via \figureref{figure:quantiles}. As discussed previously, interventions are determined by ranking individuals according to their predicted score and then intervening on those in a bottom fraction. Hence, we evaluate the performance of these targeting strategies by examining the characteristics of students in the tail end of risk scores generated by each model.

We see that for students predicted to be in the bottom 11\% of individual risk scores on the test set, the mean graduation rate is 61\%. On the other hand, for students in the bottom 11\% of environmentally generated risk scores, the average graduation rate is 68\%.
In other words, at the same level of resource constraint, the environmental predictor targets a group of students only slightly more likely to graduate than those identified by the individualized risk scores. 
Moreover, no student in the bottom 11\% of environmental risk scores has a predicted \emph{individualized} risk score higher than .85.\footnote{For contrast, less than .05\% of students lie in the top 11\% of environmental risk scores and are predicted to have lower than an 85\% chance of graduating by the individual model. Said otherwise, good schools virtually have no struggling students.} If we take the individualized predictor as ground truth, this again shows that none of the students identified as high risk by the environmental predictor have graduation likelihoods close to 1. 

In light of the estimated effect sizes of interventions, intervening on the students identified as high-risk by either targeting strategy would be a perfectly efficient use of resources and have the same aggregate effect on the overall graduation rate in Wisconsin. For whatever treatment effect the hypothetical intervention has in this (-2\%, 12\%) range, there is no reason to believe its effect would be significantly better if applied to students with a 61\% likelihood of graduation versus students with an average graduation likelihood of 68\%.

\begin{figure}[t!]
\begin{center}
\includegraphics[width=.49\textwidth]{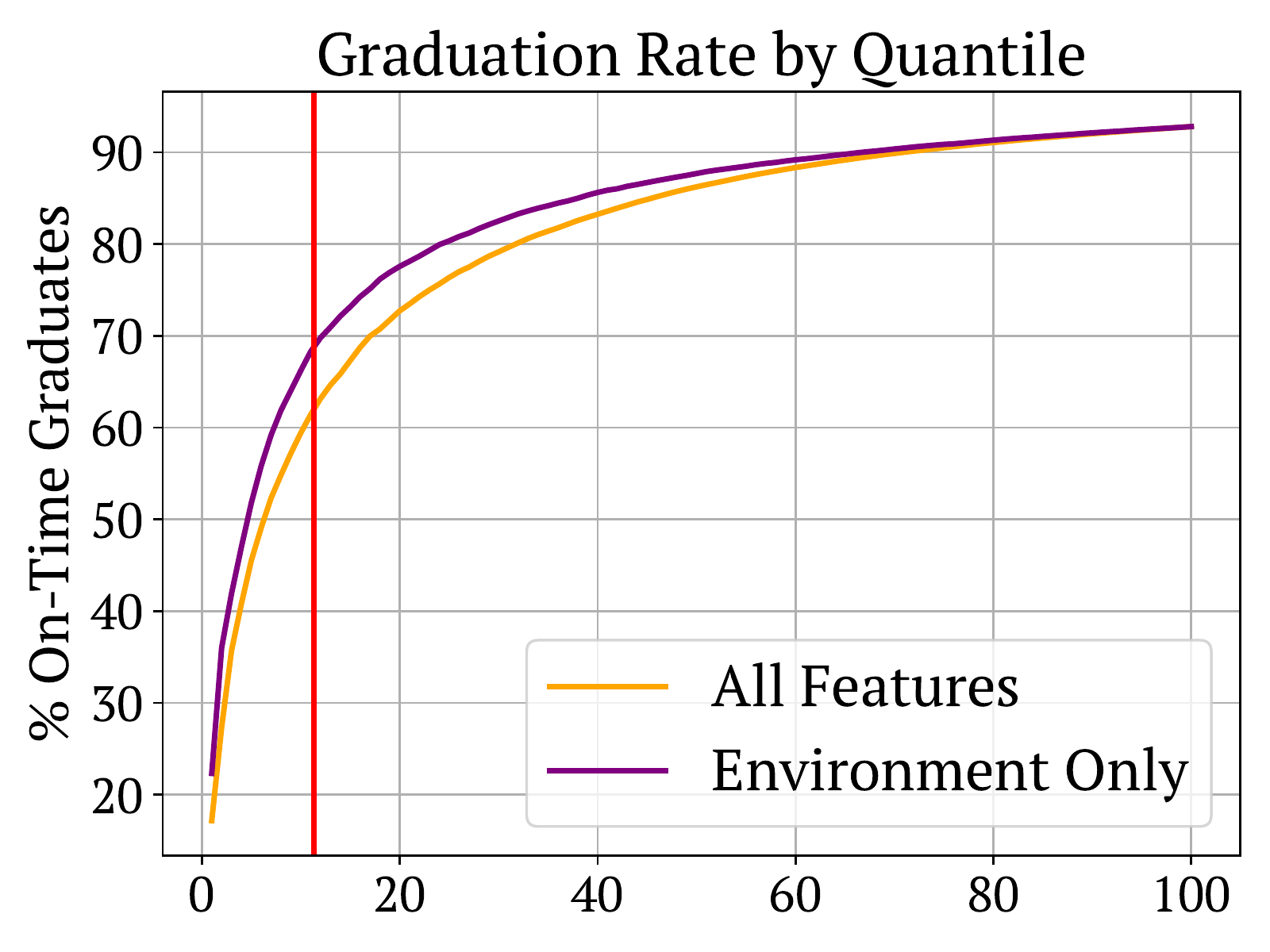}
\includegraphics[width=.49\textwidth]{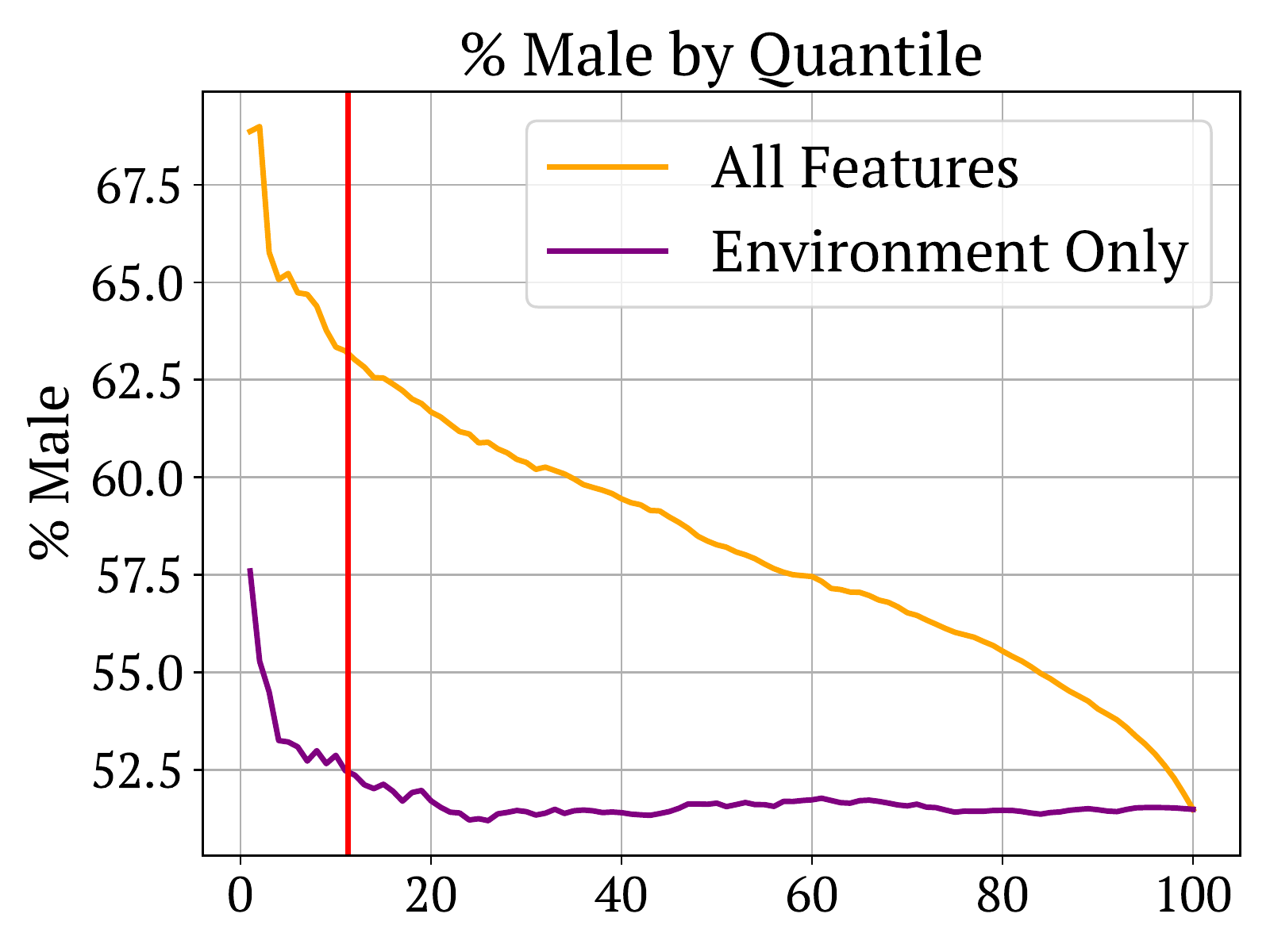}
\includegraphics[width=.49\textwidth]{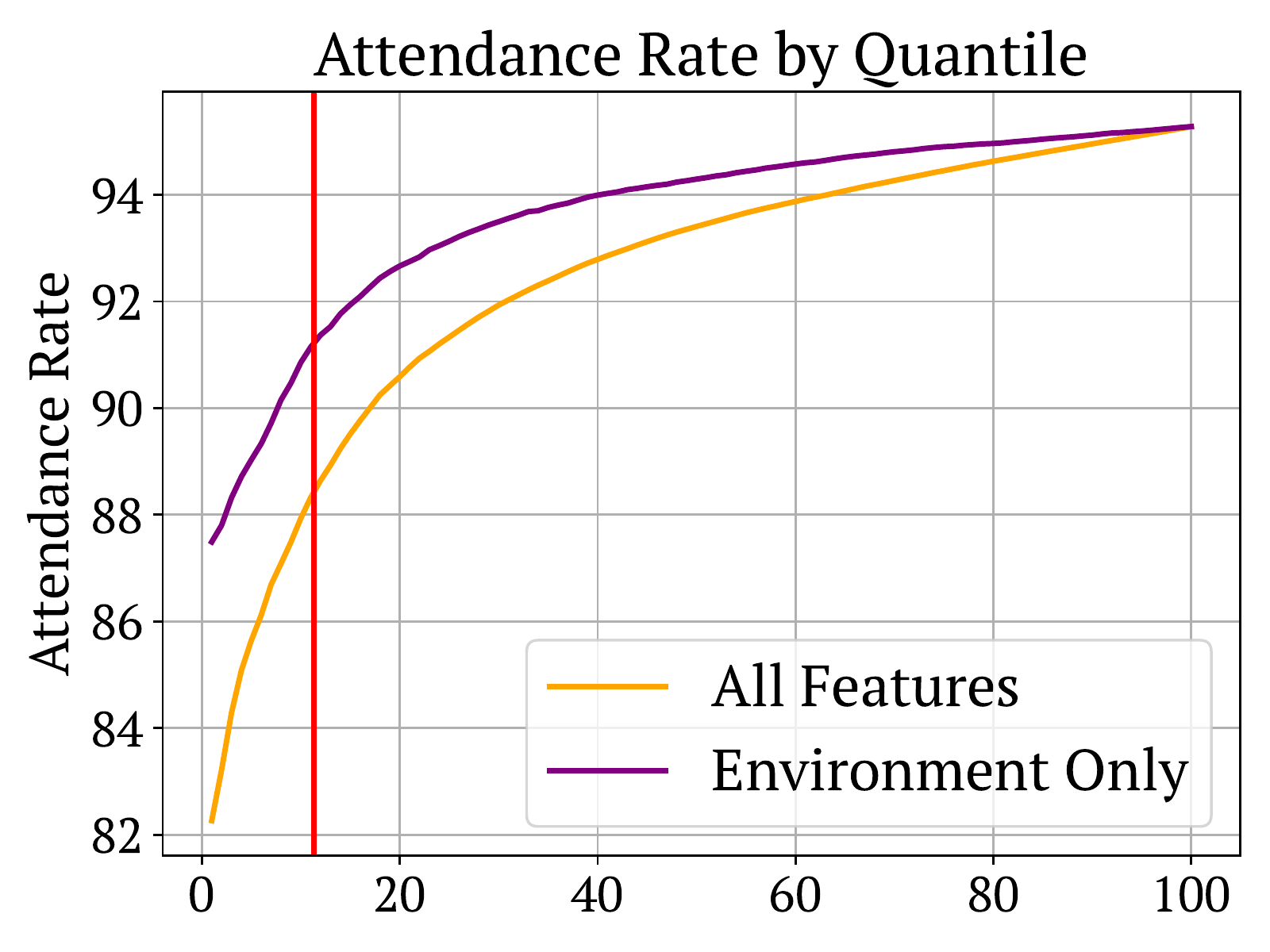}
\includegraphics[width=.49\textwidth]{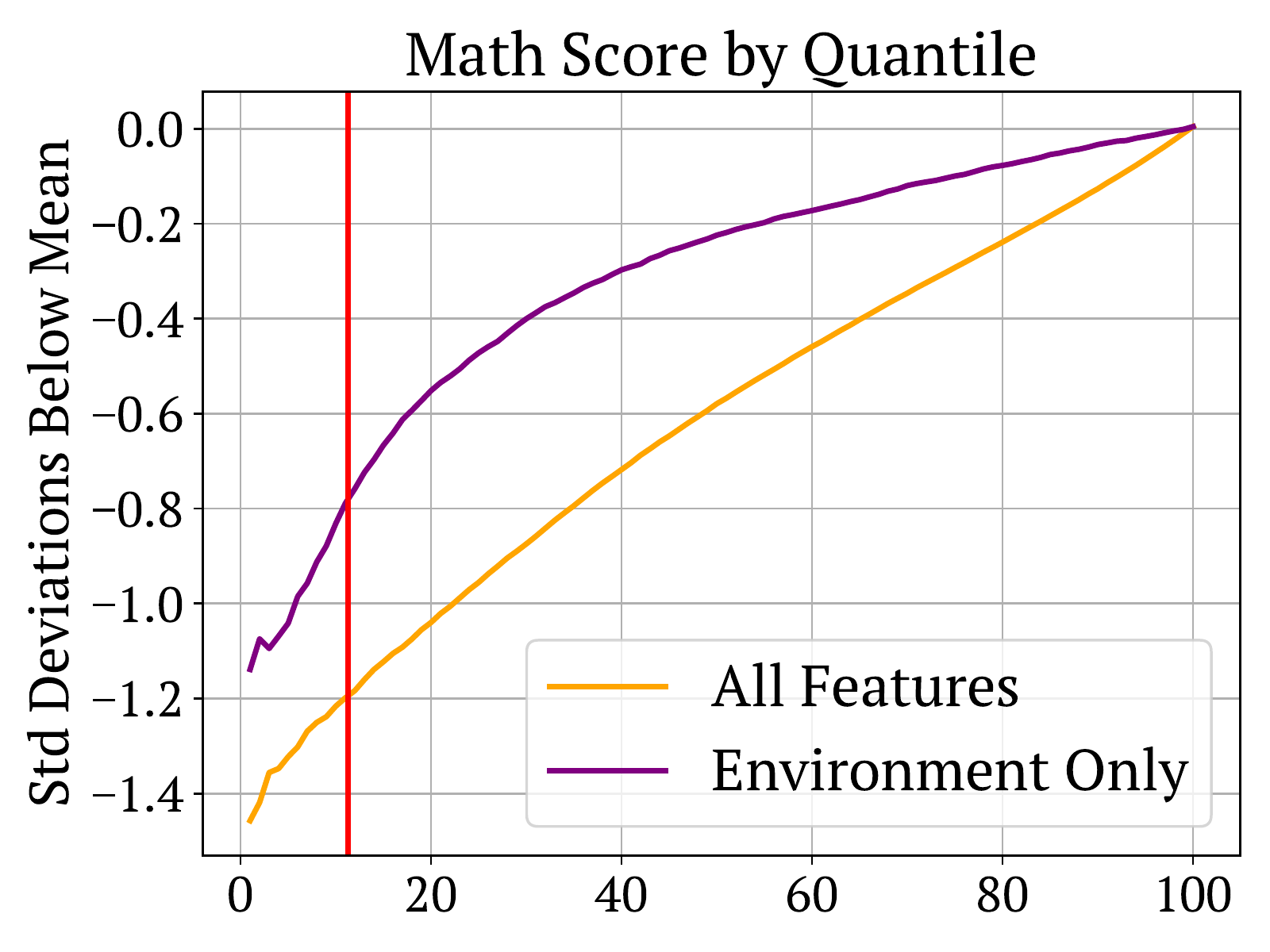}
\end{center}
\caption{Comparing targeting performance of environmental and individual predictors. 
The $x$-axis describes the quantile of the predicted risk score. The $y$-axis looks at the average value of various characteristics amongst students whose predicted risk scores lie in the bottom $x$\% of the population.
\textbf{Top Left:} Average graduation rate for students in the bottom $x$\% of predicted scores ($x$ is the quantile varies from 1 to 100). \textbf{Top Right:} Fraction of male students by quantile. \textbf{Bottom Left:} Attendance rate of students by quantile of predicted score. \textbf{Bottom right:} Average math score of students by quantile. For clarity, we measure scores in terms of standard deviations below the mean. For all four plots, we draw the vertical line at the vertical red line at $x=11$, the fraction of students predicted to be at high risk by DEWS.}
\figurelabel{figure:quantiles}
\end{figure}

While the groups of students identified as high risk by either strategy are similar in terms of their likelihood of on-time graduation. There are notable differences in other dimensions. The high risk group for the environmental predictor is 52.5\% male, while the high risk group for the individual predictor is nearly 63\% male. Allocating interventions based on the individualized predictor would to lead to a gender unequal distribution of school services that merits additional explanation. 
Furthermore, the bottom 11\% of students according to the environmental predictor, has a higher attendance rate (91\% vs. 88\%), and higher test scores than the individual predictor's cohort (average math and reading scores are approximately .7 standard deviations below the mean vs. 1.1 in the individual group). This suggests that this cohort of students is closer to meeting the graduation bar and plausibly more responsive to successful intervention.

The main drawback of environmental-based targeting is that, by assigning all students in the same district identical scores (they have the same environmental features), it systematically de-prioritizes low-performing students in high-performing districts. In our analysis, 10\% of students in the test set are not labeled high risk by the environmental predictor, yet the individual predictor assigns them probabilities of graduation of less than 90\%. Students in this group would likely benefit from interventions, but would be less likely to receive one in the school-focused system. Nevertheless, by virtue of not being in the bottom 11\% of environmental scores, these students are in overall better-resourced schools, with fewer at-risk students, where counselors and staff are more likely to have time to devote to them.


At the same time, an environmental predictor has several distinct advantages. It is substantially simpler to build, requires less fine-grained data collection, and is easier to understand. Furthermore, it widens the scope of available interventions.
gSchool-based targeting directly prioritizes schools in need of additional resources. Many interventions, such as smaller classrooms and improved facilities, do not apply to individuals but to schools. Therefore, the set of interventions we can couple with an environmental predictor is a strict superset of those that apply to individuals only. 

\section{Discussion}

Summarizing the main empirical results of our work, we find that over the past decade DEWS has provided accurate insights into the likelihood of on-time graduation for students in Wisconsin public schools. 
In addition to accurately sorting students according to their dropout risk, we find limited evidence that students highlighted by the system received interventions that improved their outcomes. 
Yet, differences in the predictive performance suggest that students from historically disadvantaged groups may have been systematically deprioritzed for additional attention.

Drawing on these empirical insights, we propose an alternative targeting mechanism to DEWS that focuses on schools and districts rather than individual students.
In the case of Wisconsin, this strategy comes with no efficiency loss relative to current iterations of EWS, yet has several distinct advantages. Namely, it addresses the two main difficulties in EWS previously identified in the literature: implementation and usage \citep{faria2017getting,balfanz2019early, mac2019efficacy}.

The primary reason why this system work is because districts are heavily segregated across socioeconomic lines. These differences make dropout easily predictable from environmental features alone. In Wisconsin,
the vast majority of students at high risk of dropping out of school are concentrated in a few, poor urban centers. 
Virtually no students in affluent, high performing districts are predicted to leave school early. 
There are five high schools (out of a total of over 500) in Wisconsin whose graduation rates are close to 70\%. This 70\% number is about the same as the graduation rate amongst the students who were assigned a high-risk prediction by the DEWS system. Furthermore, over two thirds of students in these schools are Black or Hispanic and over 90\% qualify for free or reduced lunch. Taken together, these schools generate 10\% of all students who leave high school without a diploma in the state. Raising graduation rates in these schools closer to the state average (90\%) would significantly reduce the dropout rate.

Evidently, the task of efficiently targeting educational resources to students and identifying who needs help is not a hard problem. It can be done in a variety of ways and without appealing to methods that use very fine-grained individual information about students. The primary bottleneck towards improving high school graduation rates is not targeting resources efficiently, but finding interventions, structural or otherwise, that reliably change life trajectories.



\section*{Acknowledgements}
First and foremost, we are grateful to the Wisconsin Department of Public Instruction, and in particular Erin Fath, Cark Frederick, and Justin Meyer for providing the data and models for this work and for many fruitful discussions. We would also like to thank Boaz Barak, Ellora Derenoncourt, Todd Feathers, Devin Guillory, Guanglei Hong, Michael P. Kim, Jon Kleinberg, Sendhil Mullainathan, Nathaniel Ver Steeg, Jose Eos Trinidad, and Tijana Zrnic for detailed feedback, which helped inform the course of this work. This work is partially supported by the Andrew Carnegie Fellowship Program.

\newpage
\bibliographystyle{apalike}
\bibliography{refs}

\appendix
\clearpage

\section{Investigations into DEWS System Usage}
\sectionlabel{section:usage}
\begin{figure}[t!]
\begin{center}
\includegraphics[width=.6\textwidth]{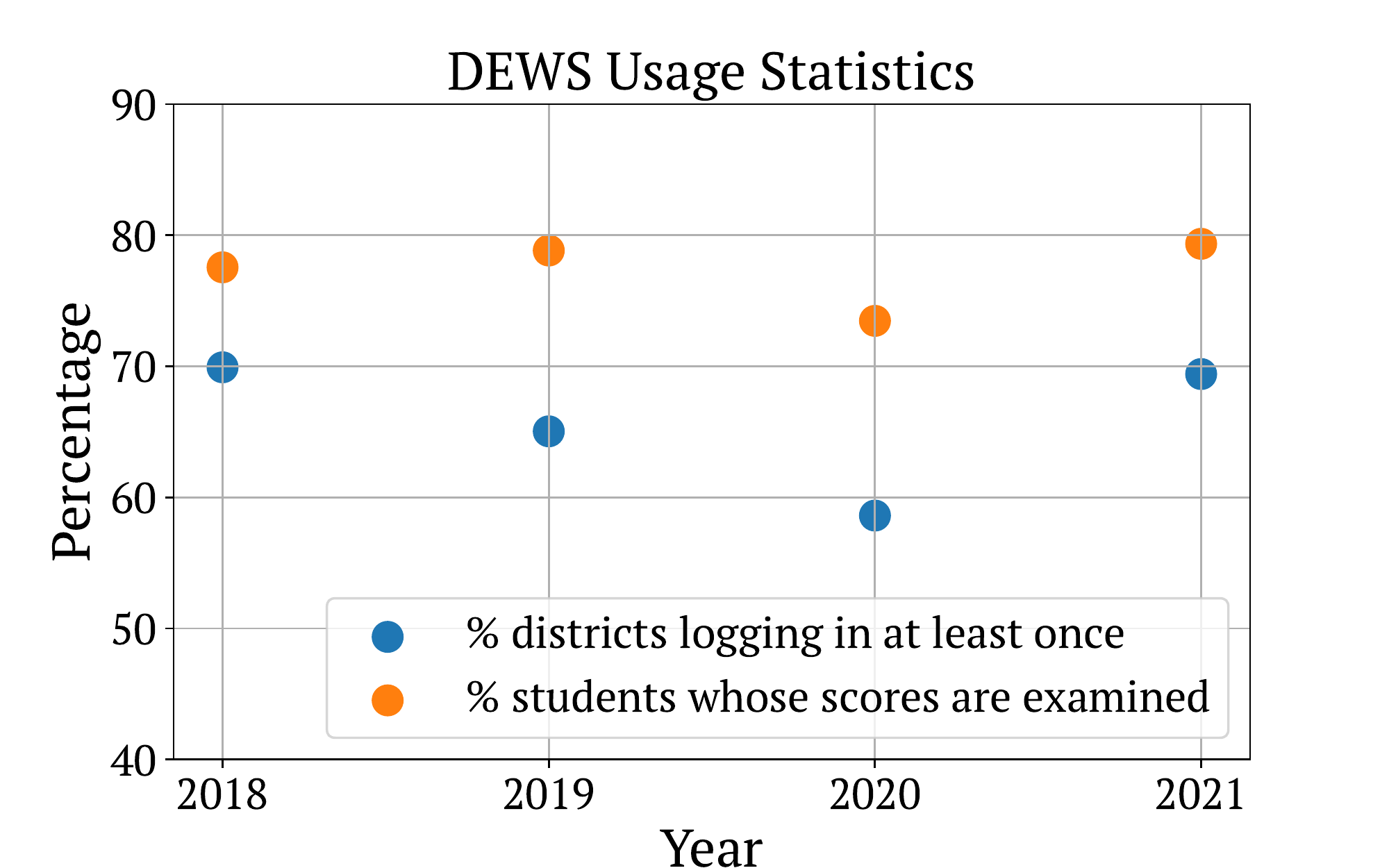}\\
\end{center}
\caption{Visit statistics describing how often staff from various public school districts in the state log onto the WISEDash platform to examine DEWS scores. In \textbf{blue}, we plot the fraction of school districts that log on at least once per year. In \textbf{orange}, we weight visits statistics by the fraction of total students in that district to arrive at a population-weighted visit statistic, measuring the fraction of students in the state whose DEWS scores are examined. See footnote for a formal description.}
\figurelabel{figure:usage}
\end{figure}

\begin{table}[b!]
\resizebox{\textwidth}{!}{\begin{tabular}{ccccc}
& Median Household Income &  \% Black & Total Population &  Graduation Rate \\
\hline
\centering
\% Years Visited  &  -.01 & .08 & .20 & -.16 
\end{tabular}}
\caption{Correlation between the fraction of years a school district visited the WISEDash portal between 2018 to 2021 and district-level socioeconomic statistics drawn from the ACS 5 year community survey.}
\tablelabel{table:correlations}
\end{table}

Predictions that are never looked at can never influence outcomes. Hence, as part of our investigations into the impacts of the DEWS program, we first verified that the system had been actively used. As discussed in the overview, the DPI issues credentials that school administrators can use to log onto a department website, WISEDash, and examine scores for students in their district. Each set of login credentials is associated with a unique district id. Using this data, we were able to examine the number of times each district in the state logs on to the DEWS platform each month for the period from 2018 to 2021. Due to software changes, data before 2018 is unfortunately not available. 

In \figureref{figure:usage}, we plot in blue the fraction of districts that log on at least once per year during the time frame in consideration. Because districts can directly download all student scores after a single yearly visit, we opt to measure usage by whether districts log on at least once, rather than the total number of visits. Using this statistic, we find that around two thirds of districts regularly log to the WISEDash portal each year. Other than a drop during the start of the COVID-19 pandemic in 2020, there are no clear trends in usage across time. While DEWS may perhaps have seen lower usage in the period before 2018 as the system was less well-known, from this data we conclude that utilization has been relatively consistent across time.

To gain better intuition of where the system may have been most effective, we also evaluated whether there are any major differences between districts that regularly log on and those that do not. In the orange dots in \figureref{figure:usage}, we plot an estimate of the total fraction of students whose DEWS scores are examined. More specifically, we weight the indicator variable of a school district logging onto the DEWS platform by the fraction of the state’s student population that is in that district.\footnote{If we let $d_{i,j}$ denote the indicator variable for district $i$ logging on during year $j$ and $n_j$ denote the the number of students in district $j$ (which is roughly constant across years), in orange we plot $(\sum_i d_{i,j} n_i) / \sum_i n_i$ across all years $j$.}
Because we cannot disambiguate between users from the same district but different schools, this statistic can  overstate the true fraction of students whose DEWS scores are acted upon. This bias is relatively minor, however, since most districts have just one public school in them.  With this caveat in mind, our analysis shows that the larger districts use DEWS more often. This observation is consistent with the fact that most of Wisconsin consists of small rural districts where teachers get to know students personally. Schools in these areas have less use for a system like DEWS. 

Furthermore, using data from the American Community Survey, in \tableref{table:correlations} we look at the correlation between the fraction of years for which a district logs on at least once and other socioeconomic information from that district. From this, we see that districts with lower graduation rates and higher poverty indices tend to log on most often. Summarizing, the DEWS system has been actively examined by the majority of public schools in Wisconsin. Furthermore, districts with the highest need also tend to have the highest usage.

\section{Description of System Features}
\sectionlabel{section:features}

Her, we describe the entire set of features we use in our experiments and analysis of the DEWS system. We divide these features according to the source they are drawn from: DEWS, the 5-year American Community Survey, and the National Center for Educational Statistics.

\newpage

\subsection{DEWS Features}
\sectionlabel{section:dews_features}
Below, we list all features included in the DEWS system. In addition to providing a short description, we indicate the relevant partition it belongs to for the comparisons between different feature sets discussed in \sectionref{section:environment}. 

\renewcommand{\arraystretch}{1.2}
\begin{center}
\begin{tabularx}{\textwidth}{|l|l|X|}
\hline
  \textbf{Feature} & \textbf{Partition} & \textbf{Description} \\
\hline
Gender & Individual & Student's gender, can be male or female \\
Race & Individual & Race is coded as a combination of 7 mutually exclusive indicator variables for whether student is White, Black, Hispanic, Native American, Pacific Islander, Asian, or belongs to two or more races  \\ 
Disability Status & Individual & Indicator for whether student has been identified as having a disability. See \href{https://dpi.wi.gov/sped/program}{DPI website} for background on disability classifications. Some classifications such as the Emotional Behavioral Disability, are strongly correlated with race and socioeconomic status \\ 
English Learner Status & Individual & Indicator of whether English is the student's native language. If the student is a non-native speaker, there are additional features describing whether English skills are low, moderate, or high \\ 
Free or Reduced Lunch Status &  Individual & Indicator variable for whether the student's household income is above or below a certain threshold based off of the federal poverty line. This threshold is uniform across the state \\
Retained & Individual & Indicator for whether a student has failed a grade in the past \\
Student Attendance Rate & Individual & Fraction of days in school \\
\hline
\end{tabularx}
\end{center}

\newpage
\begin{center}
\begin{tabularx}{\textwidth}{|l|l|X|}
\hline
  \textbf{Feature} & \textbf{Partition} & \textbf{Description} \\
\hline
Enrolled Days & Individual & Number of days a student has been enrolled in school over the last year \\ 
Reading Score & Individual & Reading score on state-wide standardized exam \\ 
Math Score & Individual & Math score on state-wide standardized exam  \\ 
Full Academic Year - District & Individual & Indicator variable for whether student has been in the current school district for the entire academic year \\
Full Academic Year - School &  Individual &   Indicator variable for whether student has been in the current school for the entire academic year \\ Disciplinary Incidents Count & Individual & Number of disciplinary incidents over the previous year \\
Days Removed &  Individual & Number of days suspended from school \\  
Removal Type &  Individual & Indicator for whether student was expelled or just suspended from school \\
Disciplinary Descriptors & Individual & Separate indicator for whether the disciplinary incident was assault, drug related, involving a weapon, or other \\ 
School Count &  Individual & Number of unique schools enrolled in during the last year \\  
 District Count & Individual & Number of unique school districts enrolled in during the last year \\  
Enrollment Count & Individual &  Number of enrollment spells during the last year \\
\hline
\end{tabularx}
\end{center}
\newpage

\begin{center}
\begin{tabularx}{\textwidth}{|l|l|X|}
\hline
  \textbf{Feature} & \textbf{Partition} & \textbf{Description} \\
\hline
Cohort Reading Scores & Environmental & Mean and standard deviation of student's school cohort reading scores on state exams \\ 
Cohort Math Scores  & Environmental &  Mean and standard deviation of student's school cohort math scores on state exams \\ 
Cohort Size &  Environmental & Number of students in cohort \\ 
Cohort Suspended & Environmental & Number of  peers in cohort who have at least one suspension \\ 
\% of Cohort with a Disability  & Environmental &  Percentage of peers in student's cohort that have a disability  \\ 
\% of Cohort FRL. & Environmental & Fraction of cohort qualified for free or reduced lunch \\
\% of Cohort Non-White & Environmental & Fraction of student's cohort that is non-White  \\
Cohort Attendance & Environmental & Mean and standard deviation of attendance rate for cohort \\
\hline
\end{tabularx}
\end{center}
\subsection{Features from the American Community Survey}

We downloaded this data from the 2015 5-year American Community Survey using the $\mathsf{censusdata}$ Python package. Data is aggregated at the public school district level and matched to students via district IDs provided by the department. While student features are associated with a specific school year, we perform a many to one mapping and assign all students from the same district (and across all available school years starting in 2013-14) to the same ACS data. These community level statistics are, however, stable across time hence the mismatch in years is relatively minor. 

We list the ACS variable codes and names below. The interested reader can visit the \href{https://www.census.gov/data/developers/data-sets/acs-5year.html}{ACS website} for a more comprehensive description regarding how these variables are defined and measured. All of these features fall into the environmental partition.

\begin{center}
\begin{tabularx}{\textwidth}{|l|X|}
\hline
Variable Code & Name\\
\hline
	DP02\_0006P &  Percent of families with male householder, no wife present \\ 
	DP02\_0008PE & Percent of families with female householder, no husband present \\
	DP02\_0151PE & Percent of total households with a computer \\
	DP05\_0001E  &  Estimate of total population \\
	DP02\_0152P &  Percent of total households with a broadband Internet subscription \\
	DP02\_0059PE & Percent of population 25 years or older with less than a 9th grade education \\ 
	DP02\_0060PE & Percent of population 25 years or older who completed at least 9th grade \\
	DP02\_0061PE &  Percent of population 25 years or older with a high school degree \\ 
	DP02\_0062PE &  Percent of population 25 years or older who attended some college, but did not graduate \\
	DP02\_0063PE & Percent of population 25 years or older with an associate's degree \\
	DP02\_0064PE & Percent of population 25 years or older with a bachelor's degree \\
	DP02\_0065PE & Percent of population 25 years or older with a graduate or professional degree \\
	DP02\_0066PE &  Percent of population who completed high school or higher degree \\
	DP02\_0067PE &  Percent of population who has a   bachelor's degree or higher degree \\ 
	DP02\_0079PE &  Percent of population living in the same house as a year before \\
	DP02\_0087PE &  Percent of total population that is native born \\
	DP02\_0092PE &  Percent of total population that is foreign born \\
	 DP03\_0005PE &  Unemployment rate for those 16 years of age or older \\
	 DP03\_0062E  &  Median household income (2015 inflation-adjusted dollars) \\ 
	 DP03\_0096PE &  Percent with health insurance coverage \\ 
	 DP03\_0119PE &  Percentage of families whose income in the last 12 months is below the poverty level \\ 
	 DP05\_0017E  &  Median age  \\
	 DP05\_0032PE &  Percent white \\
	 DP05\_0033PE &  Percent Black or African American \\
	 DP05\_0039PE &  Percent Asian \\ 
     DP05\_0066PE &  Percent Hispanic or Latino (of any race)   \\
\hline
\end{tabularx}
\end{center}

\newpage
\subsection{Features from the National Center for Education Statistcs}

The National Center for Education Statistics maintains an online portal called the \href{https://nces.ed.gov/ccd/elsi/}{Elementary / Secondary Information System (ELSI)} whereby one can access financial data for public school districts in the US. Using their website, we pull the following set of features, each of which is associated with particular district ID and school year. This allows us to match the statistics with the dataset of individual student records. Please see the ELSI website for a full \href{https://nces.ed.gov/ccd/elsi/glossary.aspx?app=tableGenerator&term=11019,9546,25783,25789,25784,25790,25785,25778,25787,25791,25792,25793,25794&level=District&groupby=0}{glossary of terms}. All of these features fall into the environmental partition.

\begin{center}
\begin{tabularx}{\textwidth}{|X|}
\hline
Variable Name\\
\hline 
   Total Current Expenditures: Other El-Sec Programs per Pupil\footnote{El-sec expenditures refers to total current expenditures for public elementary and secondary education that are associated with the day-to-day operations of the school district.} \\
 Total Current Expenditures: Salary  per Pupil \\
Total Revenue - Federal Sources per Pupil  \\
 Total Revenue per Pupil \\
 Total Expenditures - Capital Outlay per Pupil \\
 Total Revenue - Local Sources per Pupil \\
 Total Expenditures per Pupil \\
 Total Current Expenditures - Non El-Sec Programs per Pupil \\
 Total Current Expenditures - Instruction per Pupil \\
  Total Current Expenditures per Pupil  \\
 Total Current Expenditures - Support Services per Pupil   \\
 Total Revenue - State Sources per Pupil  \\
 Instructional Expenditures per Pupil \\
  Total Current Expenditures - Benefits per Pupil \\
  \hline
\end{tabularx}
\end{center}

\newpage

\section{Supporting Analysis: Do DEWS Predictions Lead to Better Graduation Outcomes?}

In this section of the supplementary material, we present additional analyses and robustness checks supporting the claims presented in \sectionref{sec:interventions}. 

\subsection{Regression Discontinuity Design: Robustness Checks \& Uncertainty Quantification}
\sectionlabel{section:robustness}

\paragraph{Inspecting Manipulation near Threshold} A core assumption enabling regression discontinuity designs is that treatment assignment (in our case, the predicted risk category) is essentially random near the threshold. For a small choice of bandwidth, we should therefore observe that the histogram of the running variable (the upper or lower confidence bounds) on which we regress outcomes, is approximately uniform around the cutoff point. 

As illustrated in \figureref{figure:histograms}, we find that this is indeed the case. There is no evidence of ``bunching
'' or manipulation for the population of students whose data is included in the regression analysis. This uniformity of scores is consistent with the idea that treatment (predicted DEWS label) is essentially determined by random assignment near the threshold.

\paragraph{Evaluating Covariate Balance} In addition to checking for bunching, a different way of assessing whether there is indeed natural variation around the threshold is to compare the characteristics of students that fall within the bandwidth.

More specifically, we can verify that important covariates are balanced around the threshold.
In \figureref{figure:covariates}, we plot the results of this comparison. We find that students falling above or below the $\ts$ threshold have similar characteristics: they have similar attendance rates, demographic characteristics, and environmental features, amongst others. These small differences in covariates are again consistent with the idea that there is natural variation around the treatment threshold. 

\paragraph{Sensitivity to Bandwidth Choice} In the regression analysis presented in the main body of the paper, we chose the bandwidth parameter $h$ to be .01. That is, we only included students if their upper adjusted score $u(x)$ (or respectively, lower score $\ell(x)$) where within .01 of the $\ts=.785$ threshold. As discussed previously, the bandwidth parameter must be small to ensure a consistent estimate of the the treatment effect. Yet, the exact choice ``how small'' is somewhat arbitrary. 

In \tableref{table:bandwidth}, we present the results of performing the same regression analyses, but with different bandwidth parameters. We find that halving ($h =.005$), or doubling ($h=.02$), the bandwidth parameter, leads to qualitatively similar conclusions. The causal impact of assigning students to higher risk categories is in the mid single digits, but these estimates come with significant uncertainty. 

\paragraph{Understanding Effects on Subgroups} So far, we have studied what is the treatment effect of assigning students into different risk categories \emph{on average} over the entire population of students. For the sake of completeness, we also assess whether the treatment effect of assigning students into different risk buckets differs significantly if instead of averaging over the entire population of students we restrict ourselves to looking at specific subgroups of students. If treatment effects are heterogeneous, it is in principle possible for the effects to be zero on average over the entire population, but large over particular subgroups. 

In \tableref{table:subgroup}, we present the results of running the regression discontinuity analysis where the population of students is further restricted to specific demographic groups: women, students of color (non-White), or students who qualify for free or reduced lunch.  In part due to the fact that sample sizes become smaller once we restrict the analysis to specific subgroups, we find no reason to believe that treatment effects are significantly different for these subgroups.

\begin{figure}[h!]
\begin{center}
\includegraphics[width=.45\textwidth]
{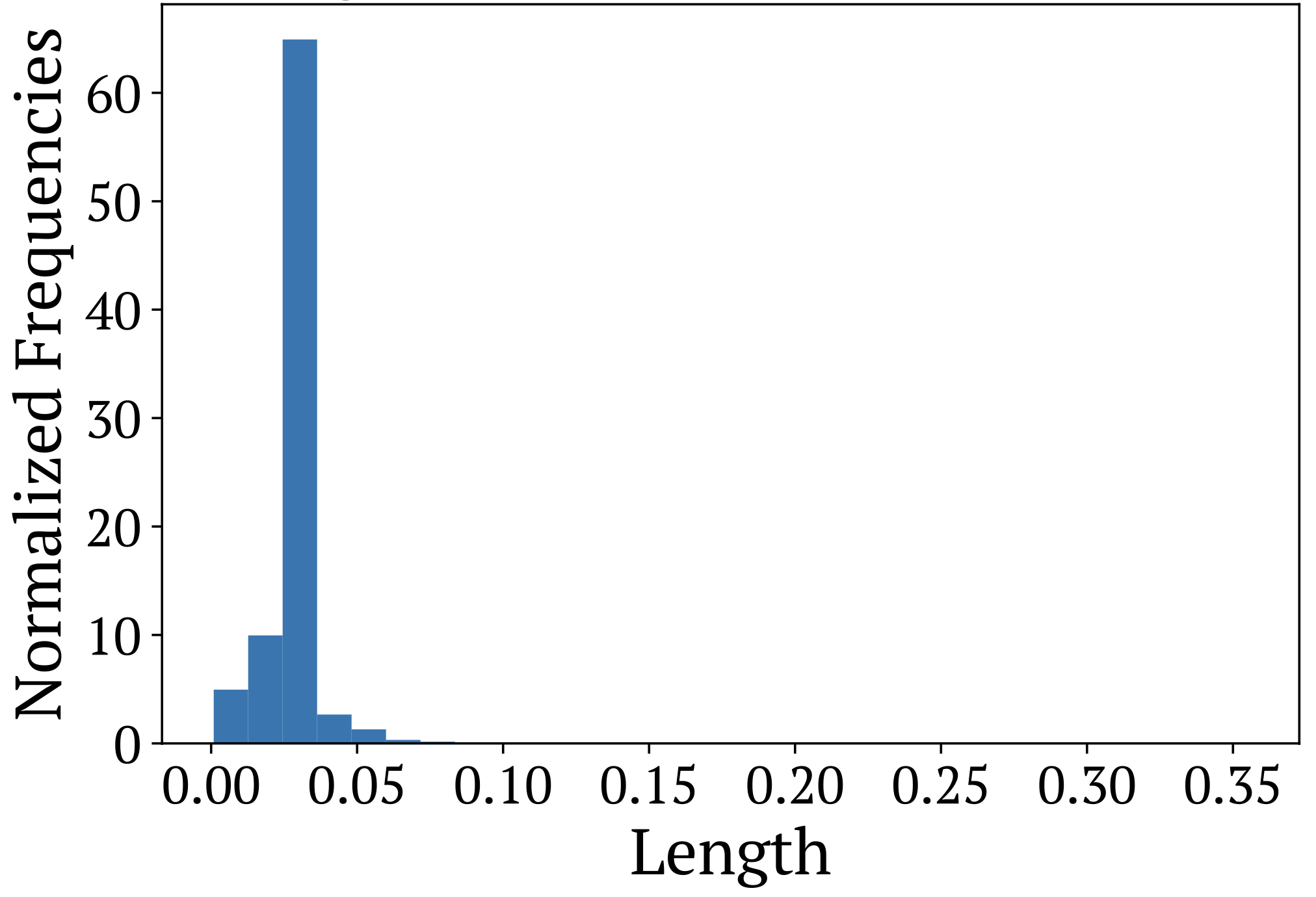}
\end{center}
\caption{Histogram of the magnitudes (length) of the error estimates $e(x)$ amongst the population of students appearing in the regression discontinuity analysis presented in \sectionref{sec:interventions}. We observe that confidence width are nearly constant, and heavily concentrate around ~.03. In particular, 95.7\% of students have error estimates that are less than .04. In other words, the upper and lower adjusted scores, that is $l(x)$ and 
 $u(x))$, are essentially equal to the DEWS score $\pm .03$.}
\figurelabel{figure:bin_width_histogram}
\end{figure}

\begin{figure}[h!]
\begin{center}
\includegraphics[width=.46\textwidth]{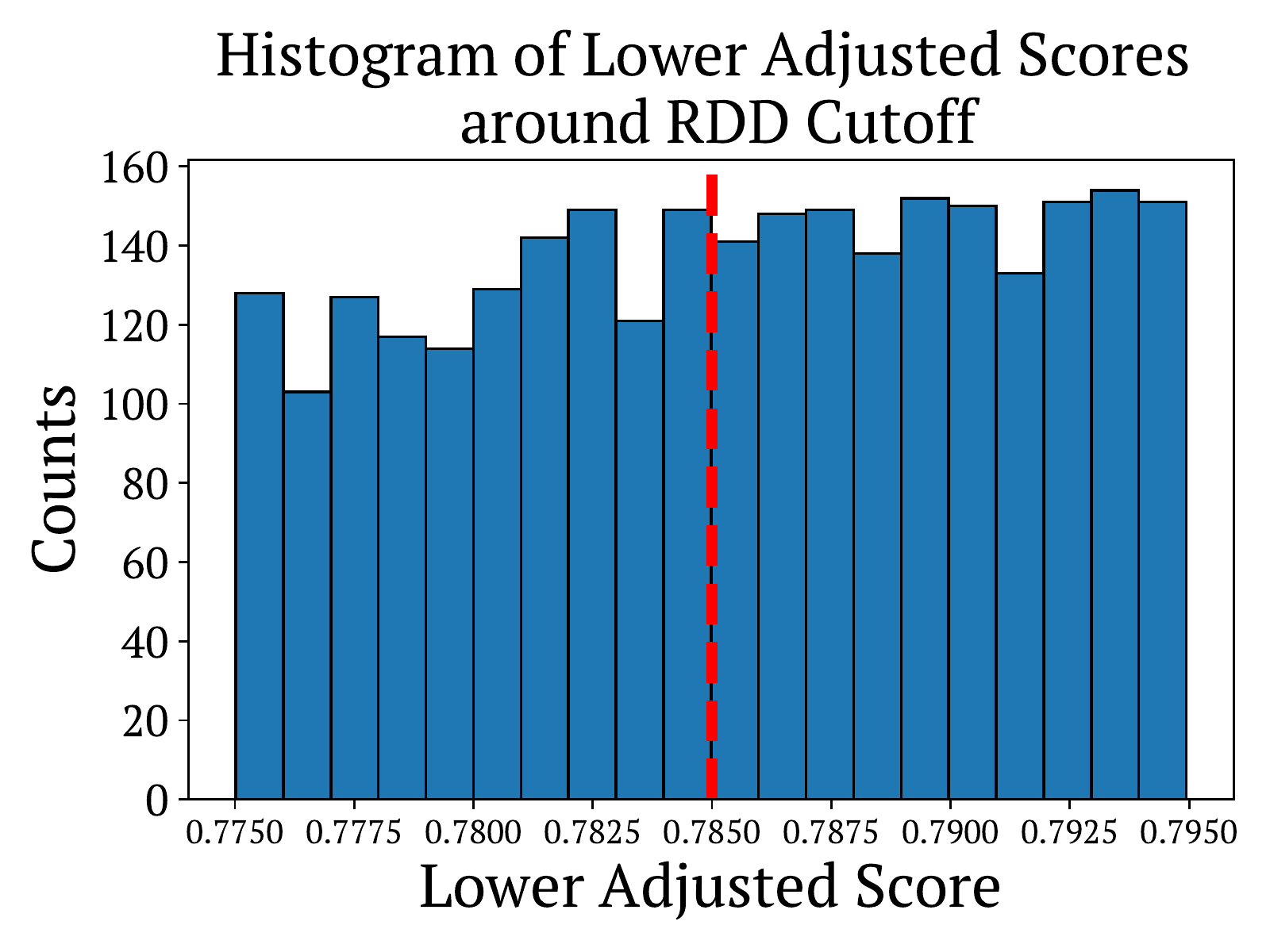}
\includegraphics[width=.46\textwidth]{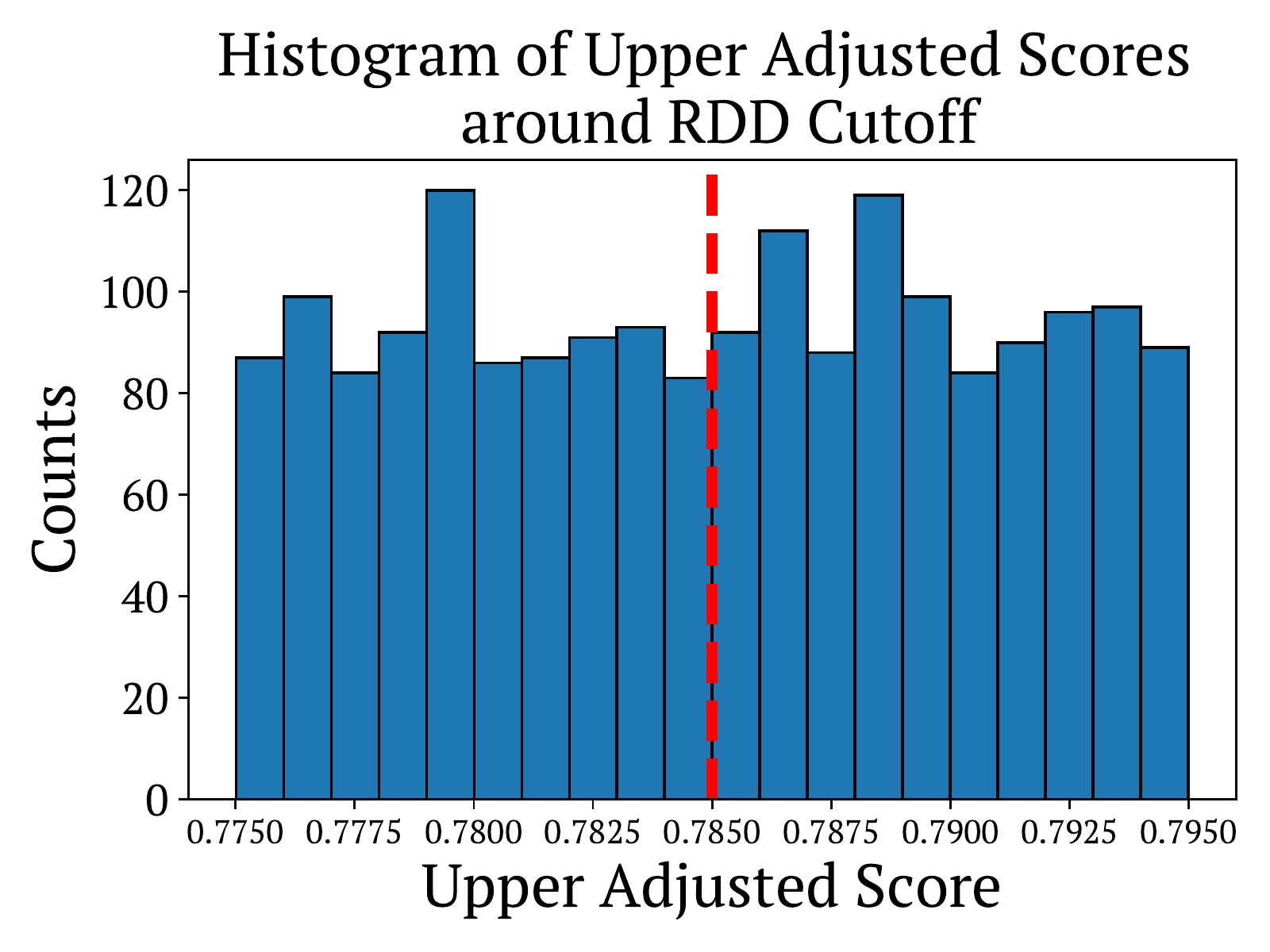}
\end{center}
\caption{Histograms of the upper and lower adjusted scores (running variables) around the threshold for both RDD analyses. \textbf{Left:} lower adjusted score, $\ell(x)$ histogram for low versus moderate comparison. \textbf{Right:} upper adjusted score, $u(x)$, histogram for moderate versus high risk comparison.}
\figurelabel{figure:histograms}
\end{figure}

\clearpage

\begin{table}[t]
\begin{center}
\resizebox{\textwidth}{!}{\begin{tabular}{cccccc}
 Causal Effect & Bandwidth $h$ &  Point Estimate & 95\% Confidence Interval &  $p$-value  & $n$ \\
\hline
\centering

  \multirow{2}{*}{ increasing risk from low to moderate} & .005 & 0.0302 & (-0.04	0.01) & 0.39 & 1390 \\
 & .02 & .0185 & (-.017, .055) & .313 & 5264 \\ 
 \multirow{2}{*}{ increasing risk from moderate to high} & .005 & 0.0574 & (-0.04,.155) & .25 & 952 \\ 
 & .02 & .0239 & (-.025, .073) & .337 & 3461 
\end{tabular}}
\end{center}
\caption{Sensitivity of regression discontinuity design to choice of bandwidth. We find that the conclusions derived from the regression analysis are largely insensitive to the choice of bandwidth $h$. Doubling or halving the bandwidth does not significantly the results presented in \tableref{table:rdd}. The point estimates are again around 5\% and 3\% for $\tmh$ and $\tml$, respectively.}
\tablelabel{table:bandwidth}
\end{table}

\begin{table}[b]
\begin{center}
\resizebox{\textwidth}{!}{\begin{tabular}{cccccc}
 Causal Effect & Subgroup &  Point Estimate & 95\% Confidence Interval &  $p$-value  & $n$ \\
\hline
\centering

  \multirow{3}{*}{ increasing risk from low to moderate} & Female & 0.001 & (-0.075, 0.078) & 0.97 & 1133 \\
 & Non White & .0124 & (-.058, .082) & .73 & 1298 \\ 
 & Free or Reduced Lunch & .0444 & (-.013,.102) & .131 & 2078 \\
 \multirow{3}{*}{ increasing risk from moderate to high} & Female & -.0007 & (-0.1,.1) & .989 & 782 \\ 
 & Non White & .0184 & (-.07, .107) & .683 & 1086 \\ 
 & Free or Reduced Lunch & .0281 & (-.05, .106) & .480 & 1476
\end{tabular}}
\end{center}
\caption{Evaluating Causal Effects of Prediction on Subgroups. The table contains the results of rerunning the regression discontinuity design analysis presented in \sectionref{sec:interventions}, but only considering students with particular features. The number of data points included in the regression is therefore smaller than that presented in \tableref{table:rdd}. The main conclusions are, however, largely identical. Treatment effects are estimated to be in the low single digits with confidence intervals containing 0.}
\tablelabel{table:subgroup}
\end{table}

\clearpage

\begin{figure}
\begin{center}
\includegraphics[width=.47\textwidth, height=3.85in]{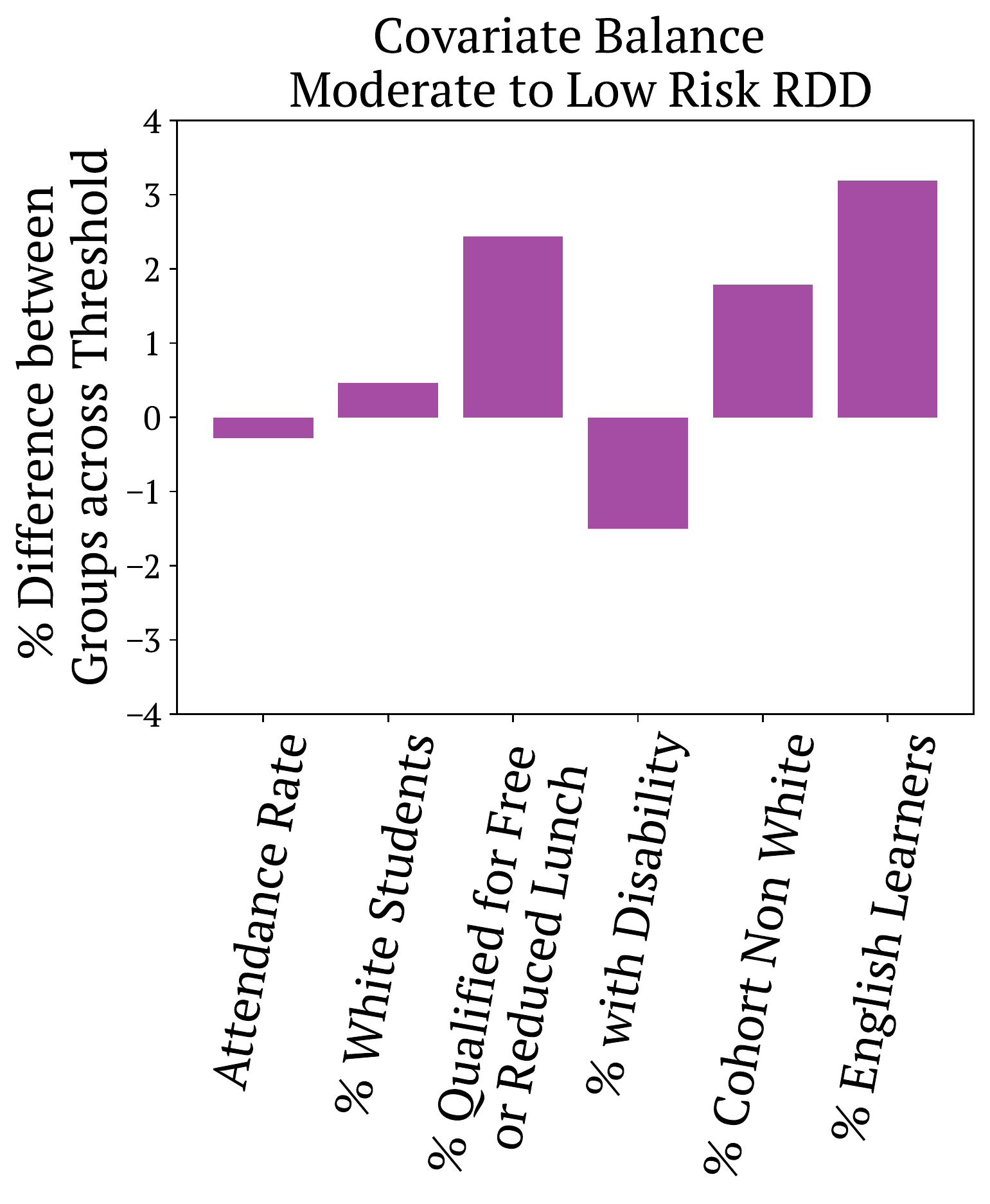}
\includegraphics[width=.46\textwidth, height=3.85in]
{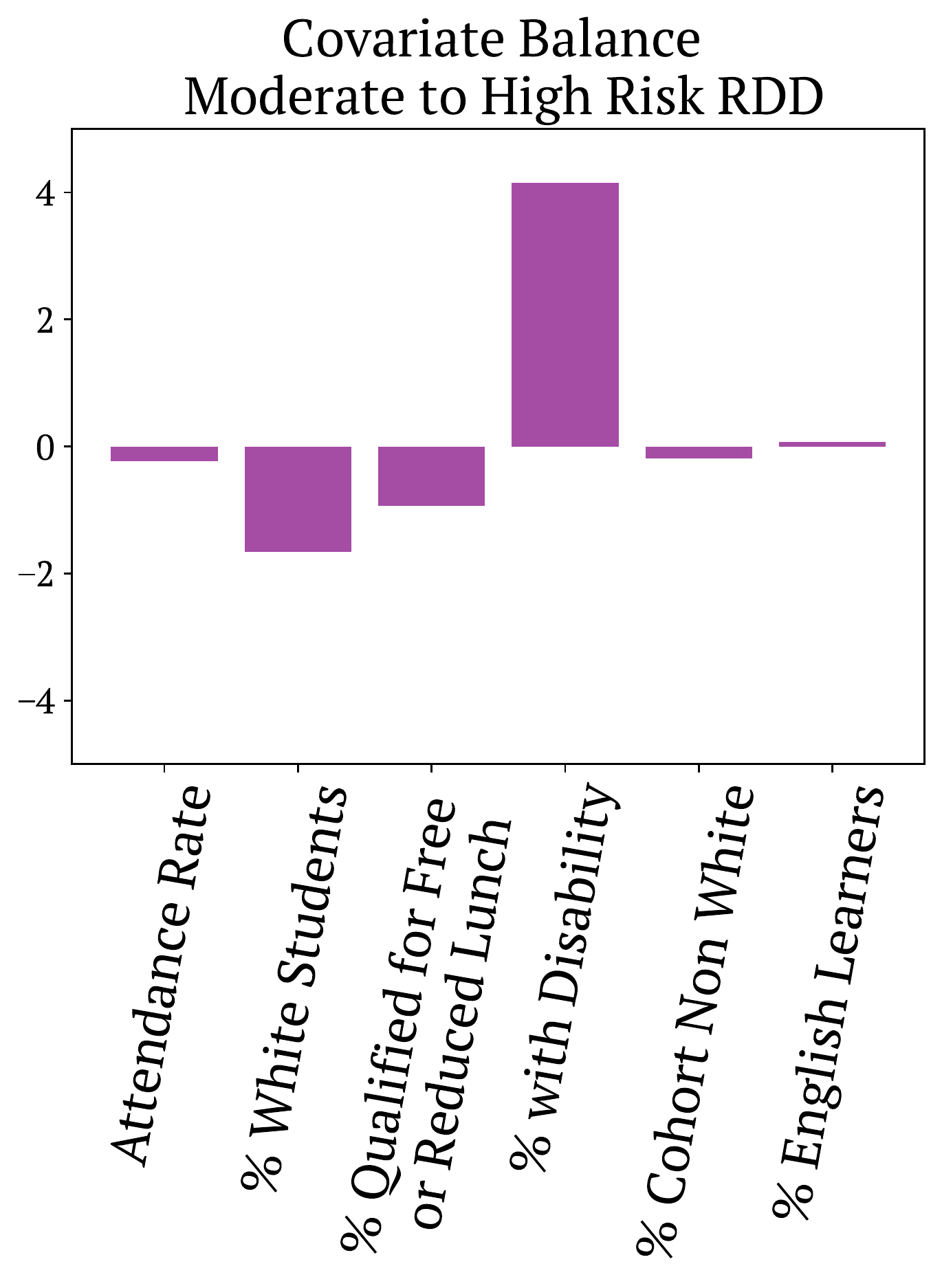}
\end{center}
\caption{Comparing the features of students on either side of the RDD threshold. \textbf{Left:} Absolute difference between the average features for students on the left and right of the threshold for the moderate to low risk treatment effect RDD analysis. \textbf{Right:} Analogous comparison for the moderate to high risk RDD.Overall, we find that students on either side of the threshold have similar features.}
\figurelabel{figure:covariates}
\end{figure}
\clearpage


%
%
%

\subsection{Assessing Statistical Independence between Predictions and Outcomes}
\sectionlabel{section:independence_tests}
Our regression discontinuity design analysis centers on the impact of the predicted DEWS label on the likelihood of graduation for a particular subset of students whose confidence intervals lie at a specific threshold. While this effect is close to zero, one might still wonder whether DEWS predictions are influencing graduation outcomes ways that may not be captured by the RDD. 

If we denote the vector of student features by $X$, the DEWS outputs (both the label and score) by $\Yhat$, and the indicator variable of on-time graduation by $Y$, a necessary and sufficient condition for predictions to impact outcomes is that predictions and outcomes are not statistically independent given the features: $\Yhat \notindep Y  \mid X$. In other words, there is a specific subset of students with features $X$ such that the choice of prediction $\Yhat$ changes the distribution over outcomes $Y$. We can represent these independence statements graphically as seen in \figureref{figure:causal_models}.

Note that if the treatment effects estimated via the RDD are nonzero, then this disproves the conditional independence statement: predictions do change outcomes. However, the converse is not true. Predictions and outcomes could be statistically dependent even if the specific treatment effect estimated by the RDD is zero.

In this subsection, we directly tackle this question of detecting whether predictions and outcomes are statistically dependent using qualitatively different assumptions than in the RDD analysis. 
To disprove a conditional independence statement, it suffices to show that a predictor which predicts $Y$ given $X$ and $\Yhat$ achieves higher predictive performance than one that just uses the vector $X$. Intuitively, higher predictive performance indicates that there is information about $Y$ in $\Yhat$ that is not fully contained in the features $X$. In order to identify dependence from this ``predicting from predictions’’ approach it is necessary for the predictions $\Yhat$ to be randomized functions of $X$. That is, we need a specific ``positivity'' condition to hold, akin to that present in randomized control trials. In the case of RDD, no such assumption is needed. Sharp regression discontinuity designs estimate treatment effects without need to appeal to randomized treatment assignment.
\footnote{See \citet{mendler2022predicting} for a more complete derivation of this ``predicting from predictions'' approach and discussion of the relevant identifiability conditions.}

While the specific DEWS model for each year generates predictions $\Yhat$ as deterministic functions of $X$, there is slight variation in models across various years. Recall that models are trained on a sliding window of the 5 most recent cohorts of students. Because training sets differ year over year, the resulting models are also different. We treat this variation between models as a plausible source of randomness that enables this statistical independence test. 
\begin{table}[b]
\begin{center}
\resizebox{\textwidth}{!}{\begin{tabular}{cccccc}
 Model &  Squared Loss & Log Loss & 0-1 Loss & AUC & $r^2$\\ 
 \hline
 Performative & 0.052, (0.050, 0.054) & .185, (.180, .190) & 0.065,	(.063, .068) & .863 & .184 \\
 Non-Performative & 0.052, (0.050, 0.054) & .185, (.180, .191) & 0.066,	(.063, .068) & .862 & .183 

\end{tabular}}
\end{center}
\caption{Results for the statistical independence test comparing the predictive performance of performative and non-performative models. Entries represent point estimates and 95\% confidence intervals derived from evaluating predictors on the test set. Confidence intervals are computed by bootstrapping. The identical performance of both models further illustrates how there is no evidence of predictions influencing outcomes according to this test.}
\tablelabel{table:independence}
\end{table}


Using the full dataset for 8th grade predictions described at the beginning of \sectionref{section:predictions}, we perform an 80/20 train-test split as in the environmental vs. individual comparison from \sectionref{section:environment}. On the training set, we generate two binary prediction models, one that predicts on-time graduation variable Y using the DEWS features $X$ and the predictions $\Yhat$ (DEWS score and label), versus one that just uses the  features $X$. Following the nomenclature from \cite{perdomo2020performative}, we refer to the model that includes DEWS outputs as additional covariates as the ``performative'' model. 
The predictions we train  are again ensembles of state-of-the-art supervised learning methods for tabular data such as gradient boosted decision trees.

If there is no relationship between outcomes and predictions beyond that which is captured by the features, then the predictive performance of these performative and non-performative predictors should be identical on the held-out test set. This is exactly the results we observe in the experiment. From \tableref{table:independence}, we see that the “performative” model that includes DEWS predictions as features, has a statistically identical performance to a model that just uses the covariates $X$ across a wide range of prediction metrics. 

However, because they rely on different assumptions, these tests do not directly contradict the conclusions found in the RDD. It is possible that the predicting from predictions approach fails simply due to a lack of  positivity: predictions are not really random from year to year. Exact randomization is hard to test for given the high dimensional nature of the covariates. Therefore, we include them for the sake of being comprehensive. They support the view that the causal estimates presented come with nontrivial uncertainty. We cannot with high confidence believe that assignment into higher risk categories reliably changes outcomes.  

\begin{figure}[b!]
 \centering
 \begin{tikzpicture}[scale=0.2]
\tikzstyle{every node}+=[inner sep=0pt]
\draw [black] (28.6,-29.7) circle (3);
\draw (28.6,-29.7) node {$X$};
\draw [black] (37.6,-21.1) circle (3);
\draw (37.6,-21.1) node {$\hat{Y}$};
\draw [black] (45.1,-29.7) circle (3);
\draw (45.1,-29.7) node {$Y$};
\draw [black] (30.77,-27.63) -- (35.43,-23.17);
\fill [black] (35.43,-23.17) -- (34.51,-23.36) -- (35.2,-24.09);
\draw [black] (31.6,-29.7) -- (42.1,-29.7);
\fill [black] (42.1,-29.7) -- (41.3,-29.2) -- (41.3,-30.2);
\end{tikzpicture}
\hspace{100pt}
\begin{tikzpicture}[scale=0.2]
\tikzstyle{every node}+=[inner sep=0pt]
\draw [black] (28.6,-29.7) circle (3);
\draw (28.6,-29.7) node {$X$};
\draw [black] (37.6,-21.1) circle (3);
\draw (37.6,-21.1) node {$\hat{Y}$};
\draw [black] (45.1,-29.7) circle (3);
\draw (45.1,-29.7) node {$Y$};
\draw [black] (30.77,-27.63) -- (35.43,-23.17);
\fill [black] (35.43,-23.17) -- (34.51,-23.36) -- (35.2,-24.09);
\draw [black] (31.6,-29.7) -- (42.1,-29.7);
\fill [black] (42.1,-29.7) -- (41.3,-29.2) -- (41.3,-30.2);
\draw [black] (39.7,-23.4) -- (43.15,-27.42);
\fill [black] (43.15,-27.42) -- (43.01,-26.49) -- (42.25,-27.14);
\end{tikzpicture}
    \caption{Causal diagrams illustrating possible relationships between features $X$, predictions $\Yhat$, and outcomes $Y$ in the DEWS system. 
    \textbf{Left:} Predictions do not influence outcomes: $\Yhat \indep Y \mid X$. \textbf{Right:} Predictions change outcomes:  $\Yhat \notindep Y \mid X$. 
    Because features are collected before outcomes $Y$ are observed, and since predictions are generated on the basis of features, there are causal arrows  $X \rightarrow \Yhat$ and $X \rightarrow  \Yhat$. 
    An effective early warning system should have data which follows the causal model on the right. }
    \figurelabel{figure:causal_models}
\end{figure}
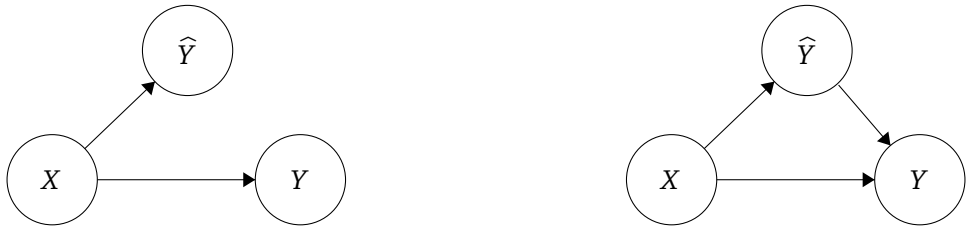

\clearpage
\section{The Strength of Environmental Predictors}
\sectionlabel{section:appendix-environment}

In \sectionref{section:environment}, we presented an alternative method of targeting resources to students that is centered on the ability to predict dropout purely from environmental variables. In this section, we develop this idea further and present a more in-depth analysis of individual versus environmental predictors in Wisconsin public schools. 

As discussed previously, our main finding is that incorporating individual-level information into a predictor that already knows coarse information describing student's academic environments only marginally improves predictive performance. To be concrete, adding individual level features improves squared loss by $\sim$ 10\% and helps identify a group of students with a 7\% lower rate of graduation. 
Given the effect sizes of available interventions (\sectionref{sec:interventions}), targeting interventions to either group would lead to an identical improvement in the statewide graduation.

The reason why these two prediction strategies yield such similar results is because school districts in Wisconsin are highly homogeneous communities. To the extent we can test in the data, there is very little diversity in terms of academic performance \emph{within} each school district. The wide range of educational outcomes comes from the fact that students \emph{between} different districts behave very differently. 
We hence observe a robust empirical pattern: Conditioning on students belonging to a specific environment (i.e the full set of environmental features), students' academic outcomes (the binary indicator of on-time graduation) are approximately uncorrelated of the full vector of individual features. Additionally knowing the vector of individual features provides little additional information regarding the likelihood of on-time graduation. 

\paragraph{Experimental Setup} We derive these insights using the exact same experimental setup as the one presented in \sectionref{section:environment}. In brief, we train two predictors --an environmental predictor and an individual predictor-- using identical learning algorithms. The only difference between both methods is the features they have access to. The individual predictor uses all the available features, while the environmental predictor is limited to using only covariates defined at the community level. Please see the previous discussion in \sectionref{section:environment} for a definition of both categories and \sectionref{section:features} for a full list.

\begin{figure}[t!]
\begin{center}
\includegraphics[width=.8\textwidth]{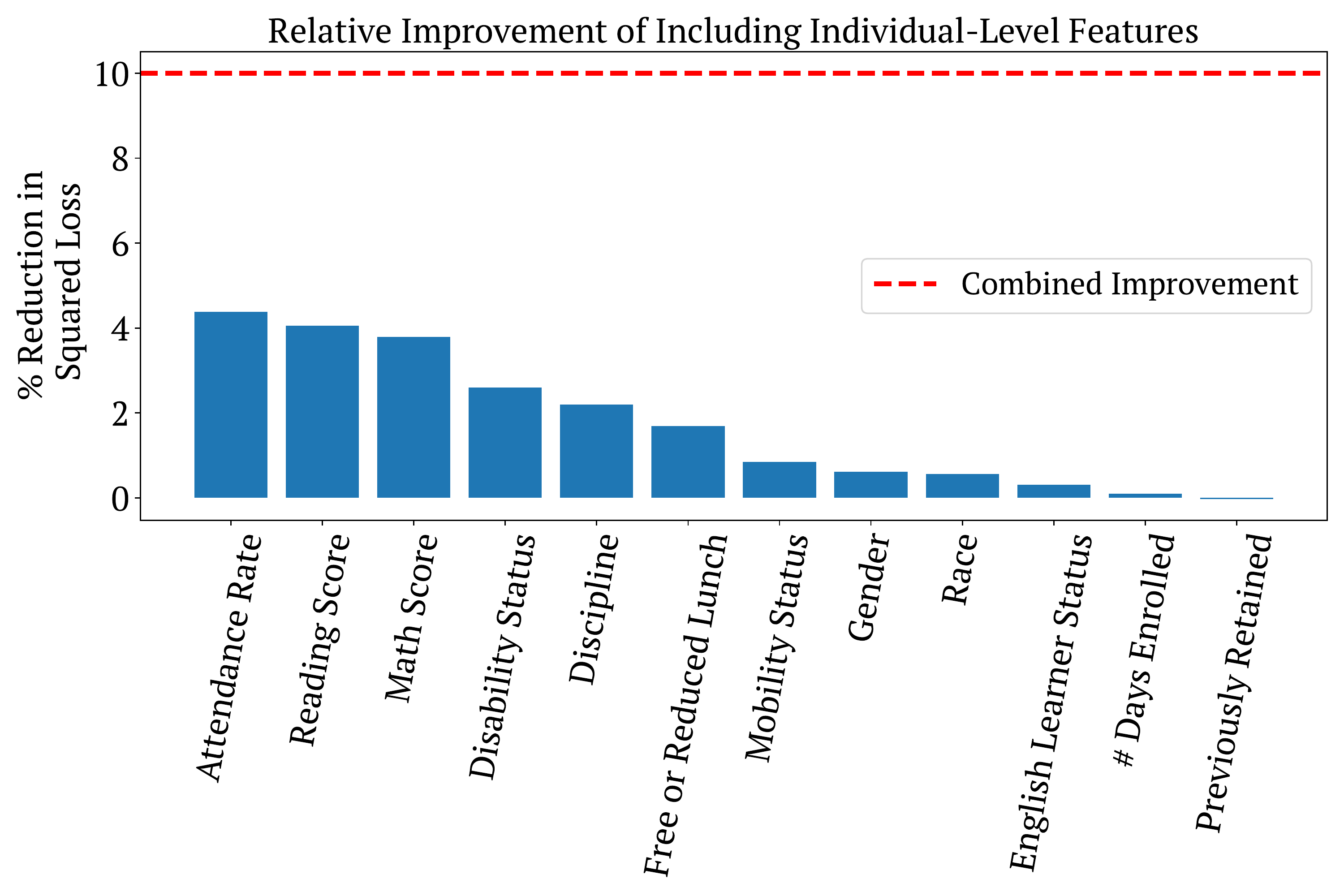}
\end{center}
\caption{Relative improvement in predictive accuracy gained by incorporating individual-level features into a predictor that only uses environmental covariates. The red dotted line indicates the percent improvement in squared loss that is achieved by including all individual features. The blue bars denote the improvement achieved by adding a single category of individual features to the model.}
\figurelabel{figure:relative_improvement}
\end{figure}

\begin{table}[b!]
\begin{center}
\begin{tabular}{ccccc}
Metric  & Squared Loss & Log Loss & 0-1 Loss & AUC \\
\hline
Absolute & .005	 & 0.03 & 0.003	 & 0.082 \\ 
Relative & 9.6\% & 14.6\% & 3.9\% & 10.2\% 
\end{tabular}
\end{center}
\caption{Relative improvement in performance achieved by including individual features on top on environmental features. For each metric, we report both the absolute improvement and the relative improvement, where relative is with respect to the performance of the environmental predictor.}
\tablelabel{table:environmental_individual}
\end{table}

\paragraph{Irrelevance of Individual Features for Prediction.} 
We begin by showing that including individual-level features in a predictive model that already uses environmental-level covariates does not lead to significantly better accuracy. Our statistical analysis  relates to a work by~\cite{hardt2022backward} who motivate a similar distinction between individual features and background features in a prediction problem, and propose statistical methods to estimate the degree to which a predictor relies on background features.

In addition to analyzing the overall value of including the complete set of individual features, we also consider the partial benefit of adding  including particular subsets of these features to an environmental-based predictor.
In particular, we repeat the same training procedure as before and produce models that predict on-time graduation using all the environmental features plus a specific individual feature.

We present the results of these experiments in \figureref{figure:relative_improvement} and \tableref{table:environmental_individual}. Adding individual features improves the predictability of on-time graduation by roughly 10\% across a number of common metrics such as log loss, squared loss, or AUC. Amongst the individual features, the single most important variable is attendance rate which improves squared loss by 4\%, followed by student's scores on standardized exams. These results are consistent with previous work on early warning systems that noted the predictive value of attendance rate on future dropout  \citep{allensworth2007matters, balfanz2007preventing}.

\begin{figure}[t!]
\begin{center}
\includegraphics[width=.49\textwidth]{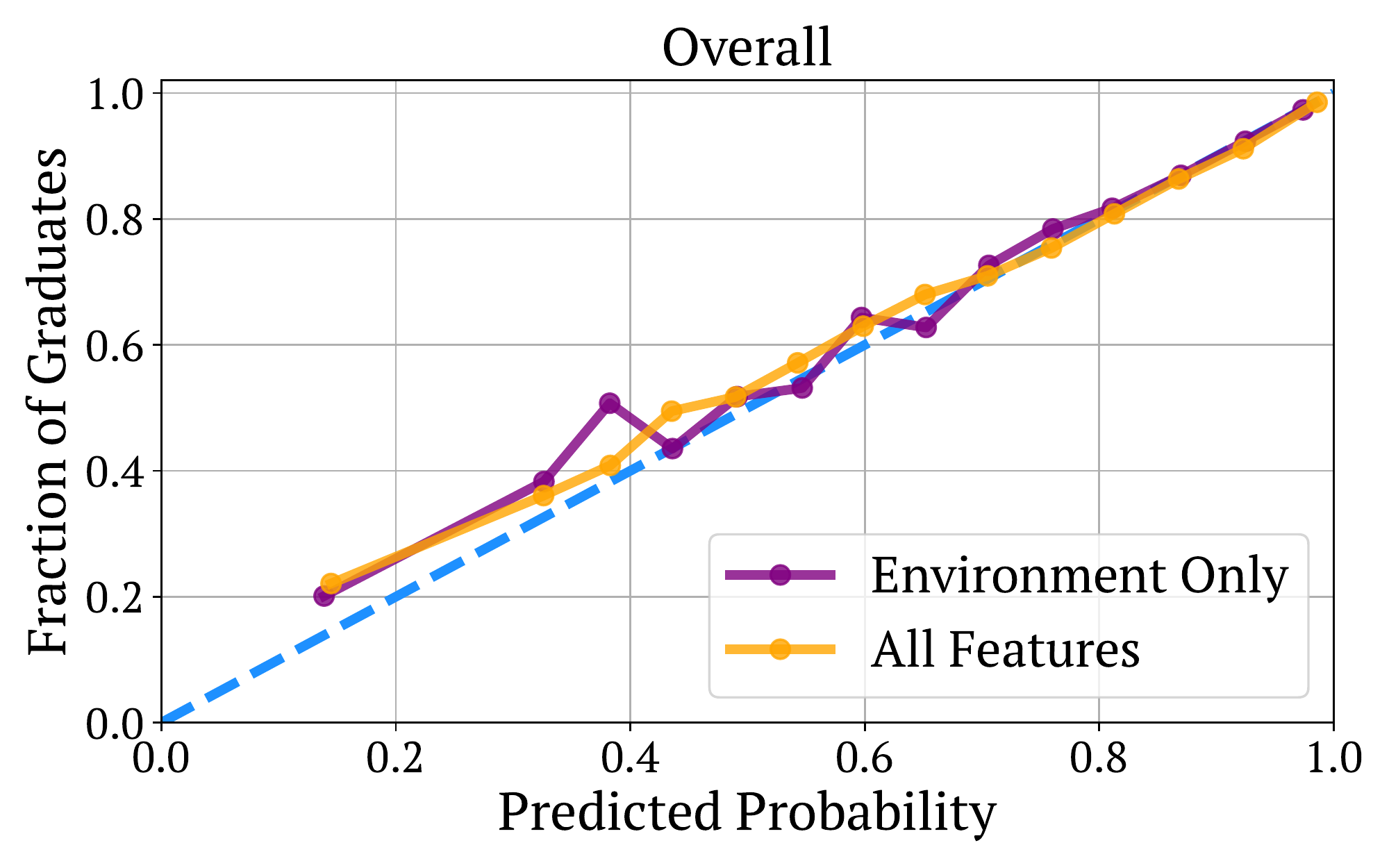}
\includegraphics[width=.49\textwidth]{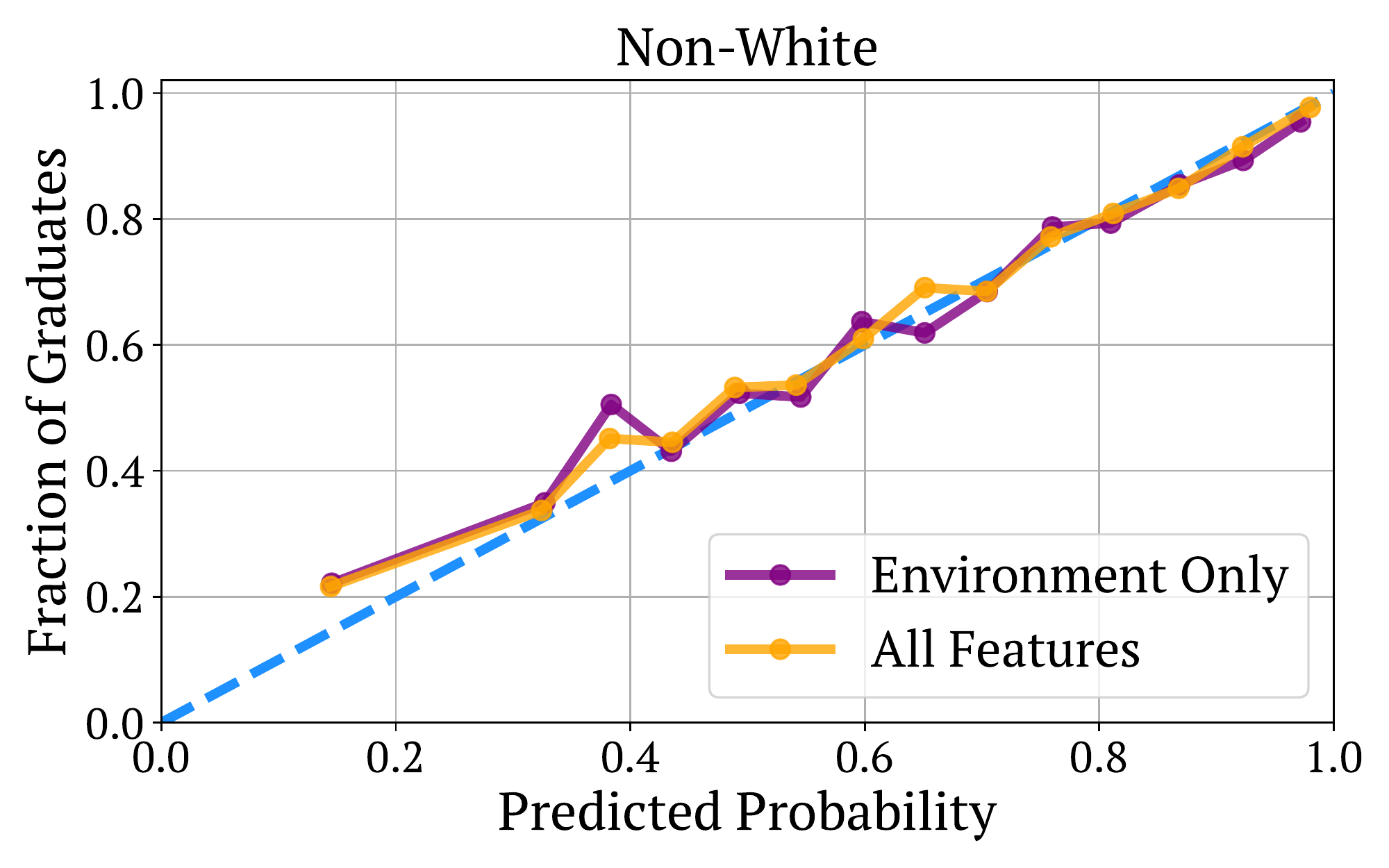}
\includegraphics[width=.49\textwidth]{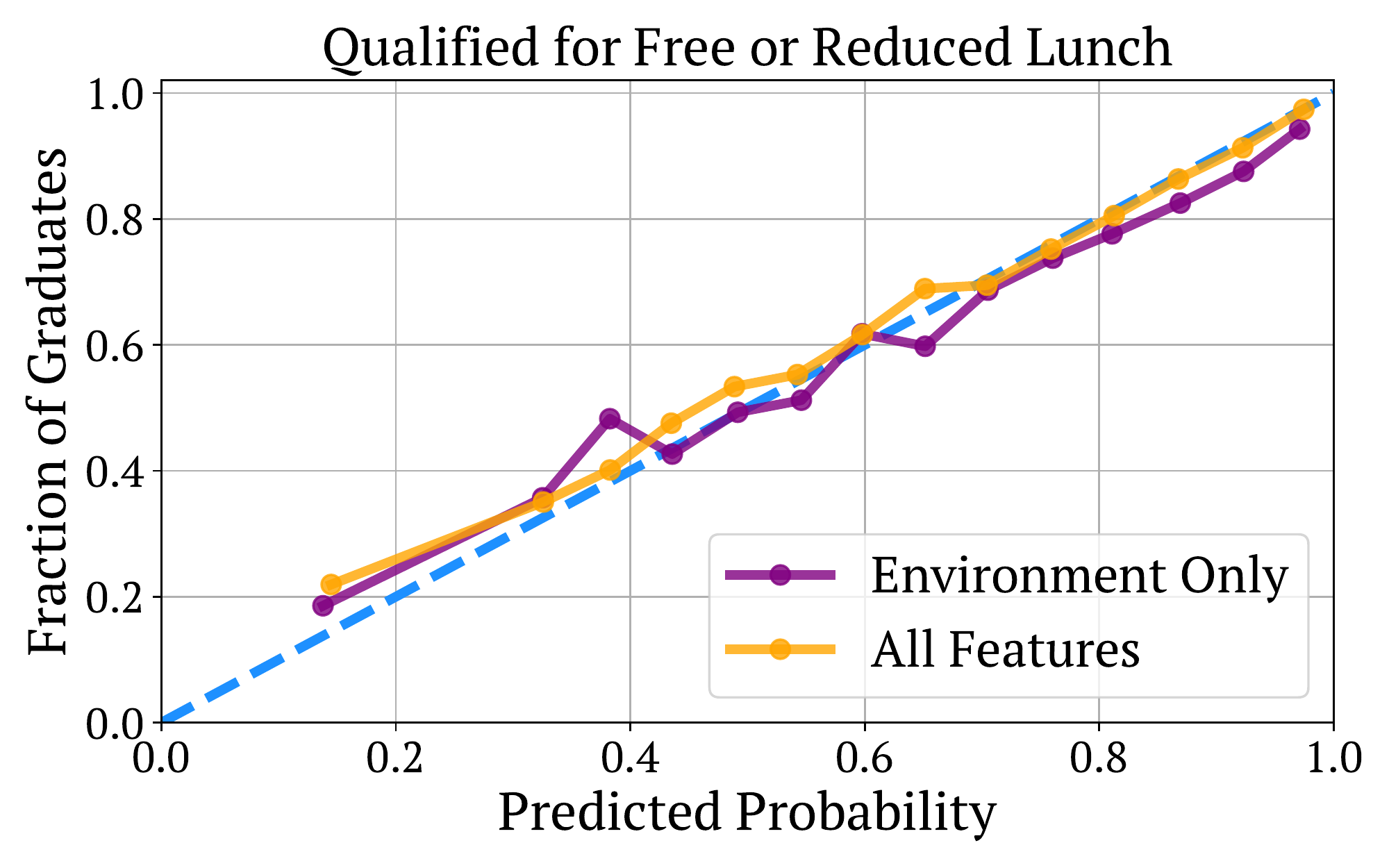}
\includegraphics[width=.49\textwidth]{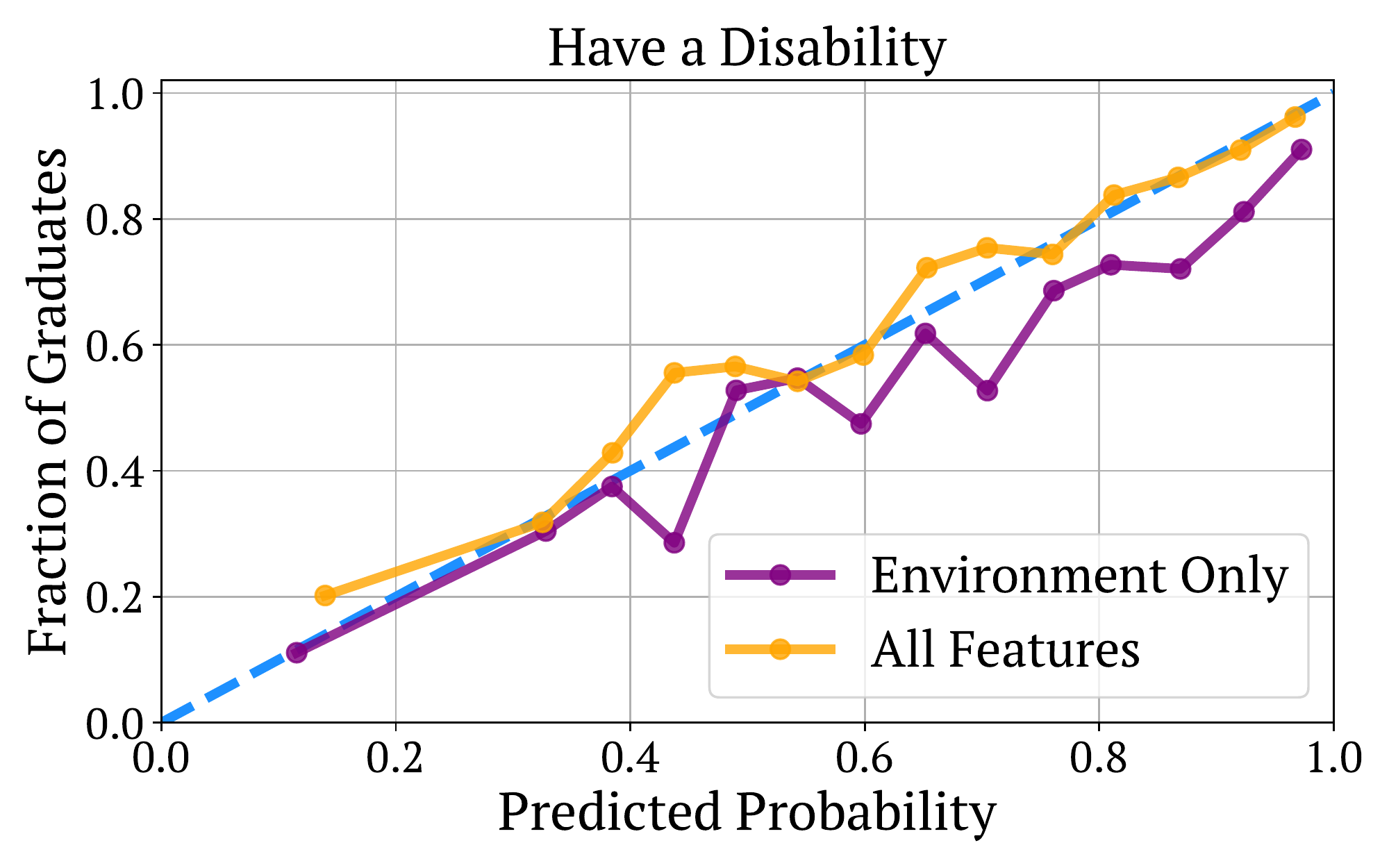}
\end{center}
\caption{Comparing calibration curves for the environmental predictor versus the full-featured predictor, amongst different individually-defined student groups. \textbf{Top Left:} comparison between white versus students of color. \textbf{Top Right:} comparison between students who have been diagnosed with a disability versus those that have not. \textbf{Bottom Left:} comparison based on qualifying for free or reduced lunch. \textbf{Bottom Right:} calibration comparison between male and female students. Overall, student of color groups tend to have lower calibration error.}
\figurelabel{figure:subgroupcalib}
\end{figure}

We examine the calibration curves of both models in \figureref{figure:subgroupcalib}.  
Both the environmental predictor and the model that uses the entire set of features generate well-calibrated predictions. The main qualitative difference between these two is that the model that uses individual features does slightly better at identifying students with low graduation rates. From looking at \figureref{figure:subgroupcalib}, the predictor that includes individual features generates slightly better predictions near the bottom left of the calibration curve. However, this difference is relatively minor since less than 2\% of students have predicted probabilities of graduation less than 40\%. See \figureref{figure:full_histogram}.

\begin{figure}
\begin{center}
\includegraphics[width=.55\textwidth]
{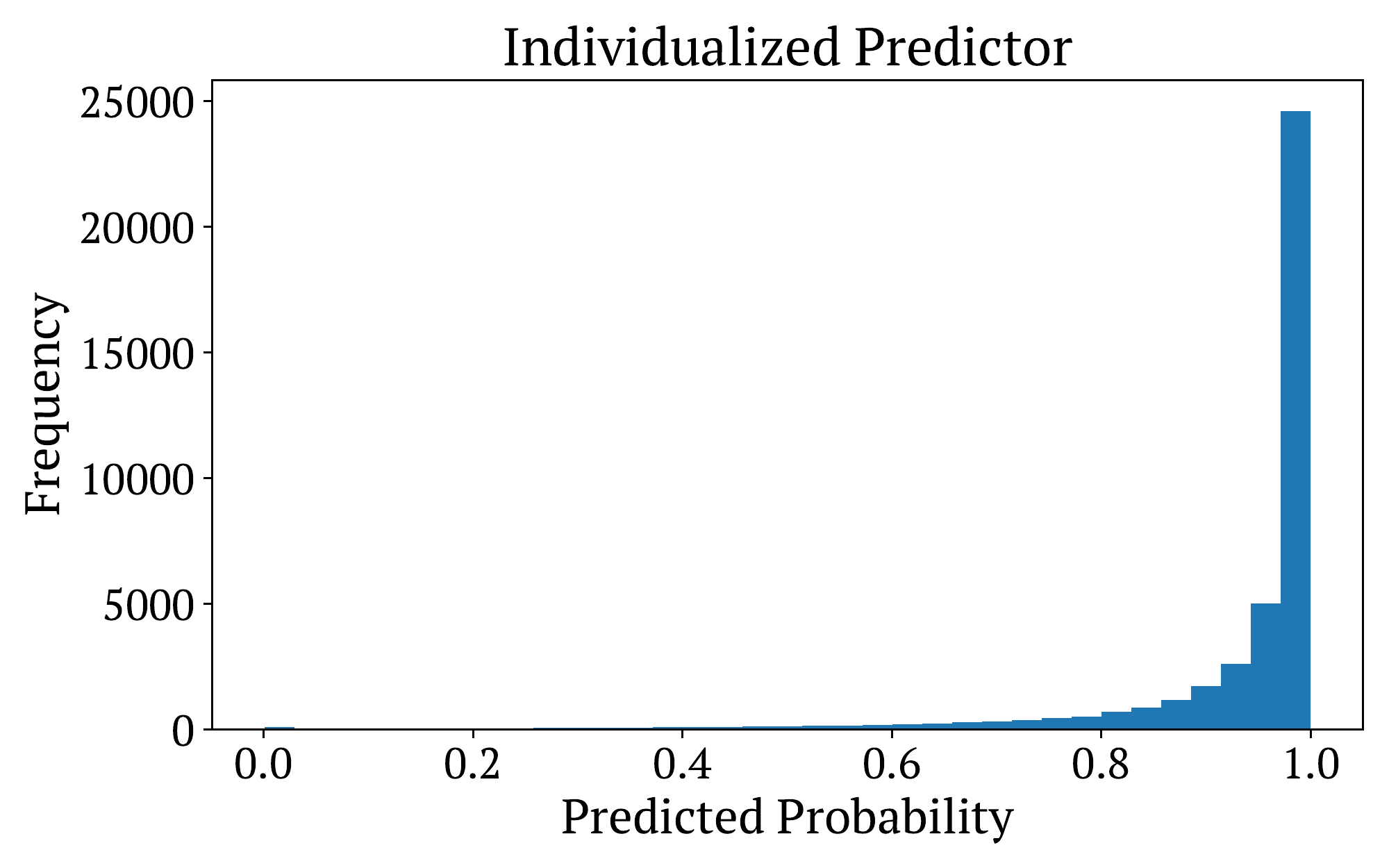}
\end{center}
\caption{Histogram of predicted probabilities on the test set generated by the individual predictor from the experiments presented in \sectionref{section:environment} that uses the \emph{complete} set of available features. The predictions from this model are highly calibrated as seen in \figureref{figure:subgroupcalib}. 10\% of students have predicted probabilities higher than .997 (that is, .997 is the 90\% quantile). 10\% of students have predicted probabilities \emph{lower} than .799 (i.e. .8 is the 10\% quantile). The vast majority of students have predicted probabilities of graduation between .9 and 1.}
\figurelabel{figure:full_histogram}
\end{figure}

Furthermore, both models remain calibrated on subgroups. In \figureref{figure:subgroupcalib}, we see that the predictions for the environmental predictor also lie close to the $y=x$ diagonal when we restrict evaluation to students of color, or students that qualify for free or reduced lunch. This is quite impressive. Note that these subgroups are defined in terms of individual-level features. Therefore, the environmental predictor does \emph{not} have access to this particular piece of important information (the indicator of group membership) when generating the prediction, yet the prediction is still well-calibrated for this group of students. Outcomes for these specific students are statistically largely determined by their environments, not these individual features. Stated otherwise, group membership (e.g., individual indicator of being non-White) is evidently highly correlated with environmental features.
\\

The reason why these two methods perform so similarly is due to the realities of socioeconomic segregation in Wisconsin public schools.

\paragraph{Same School, Similar Probability of Graduation.} 
We might imagine that within each school, there are above average students who have high probabilities of graduating on-time, a large number of people who graduate at about the school average, and a tail of struggling students that are likely to drop out. In our data, however, students in the same school have similar probabilities of graduating, regardless of their individual features.

\begin{figure}[t!]
\begin{center}
\includegraphics[width=.48\textwidth]{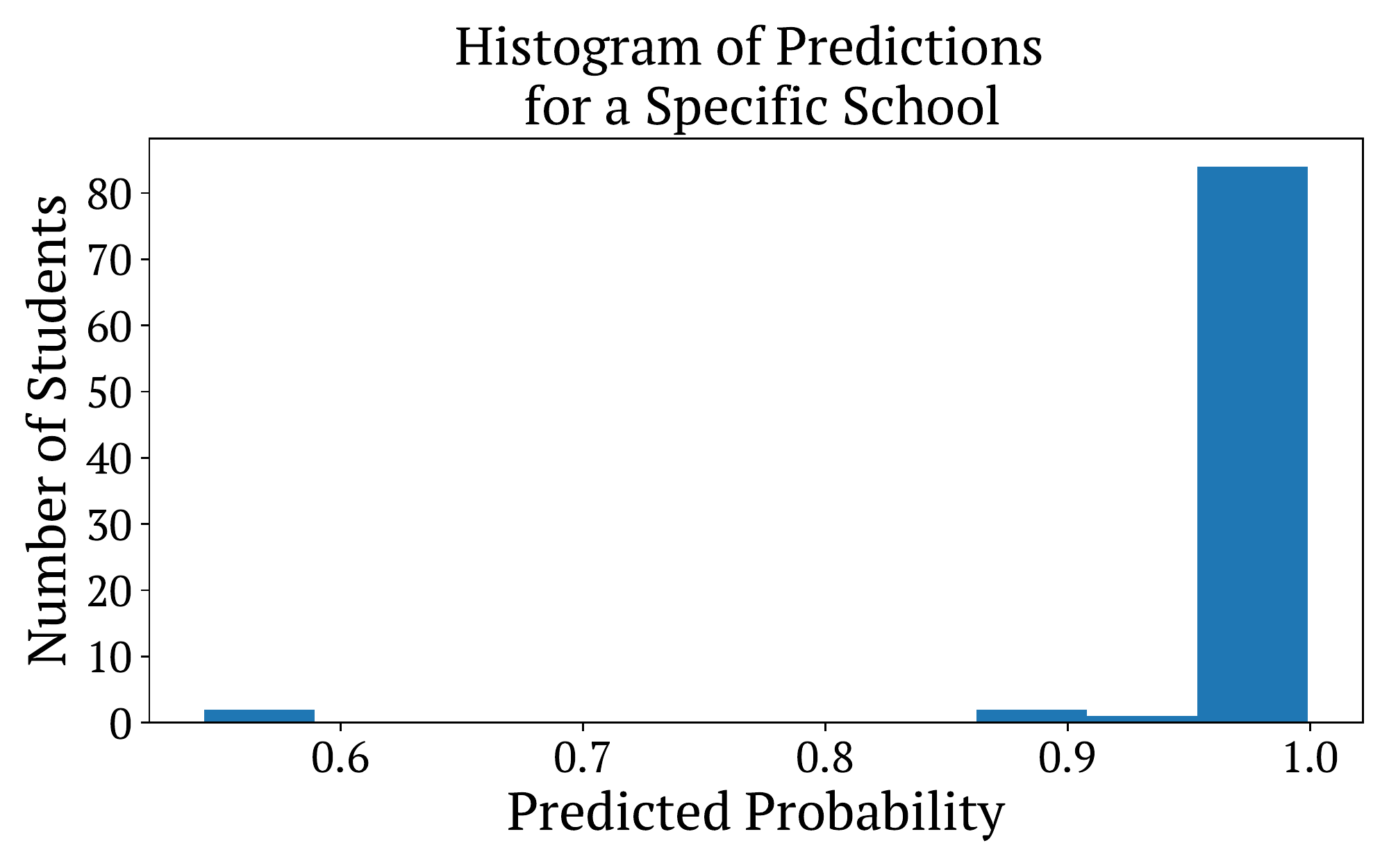}
\includegraphics[width=.48\textwidth]{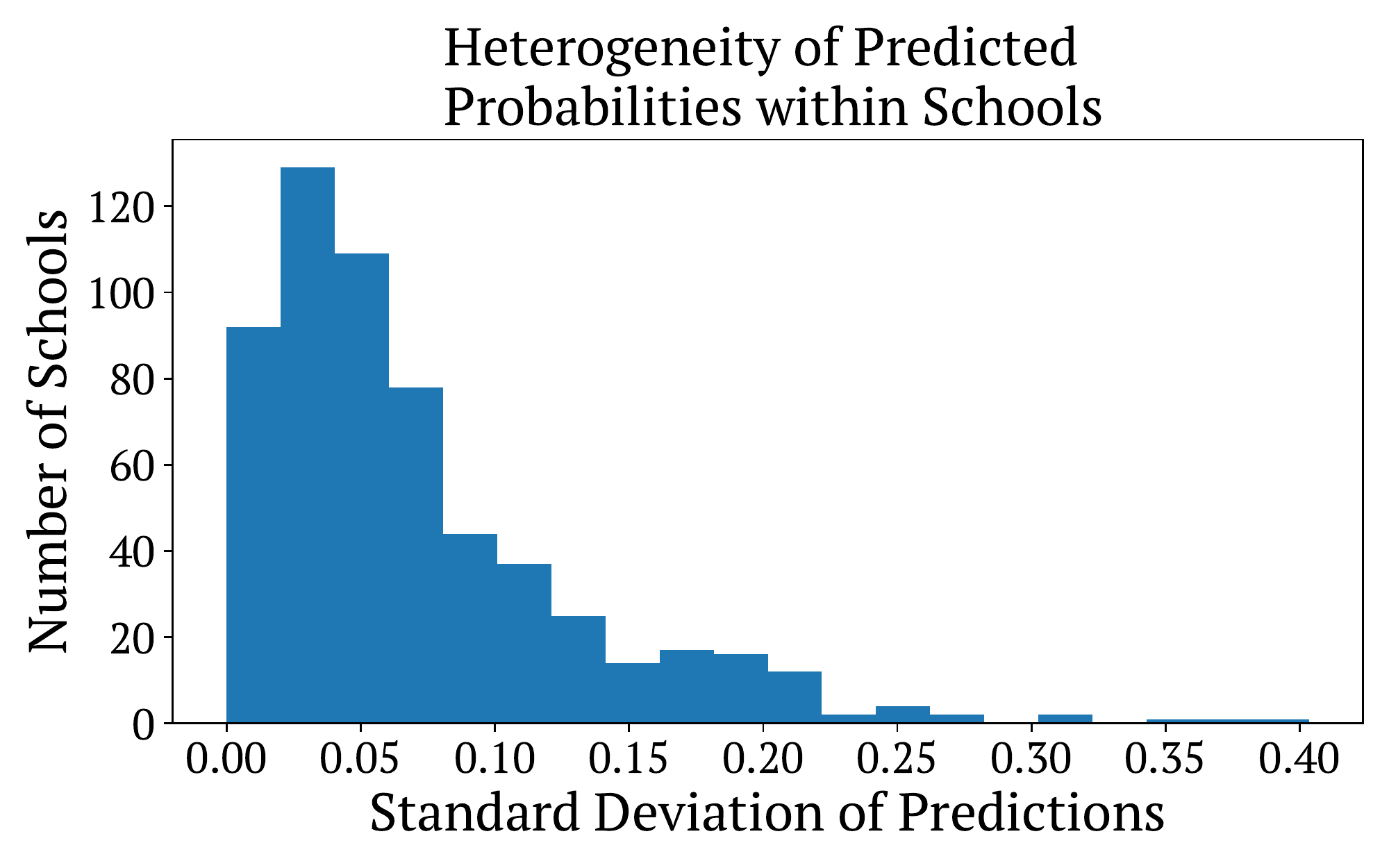}
\end{center}
\caption{\textbf{Left:} Histogram of predicted probabilities generated by the model that uses both environmental and individual features for a students belonging specific school in the held-out test set. \textbf{Right:} Histogram of the standard deviation of predicted probabilities within each school, across all schools.}
\figurelabel{figure:spread_predictions}
\end{figure}

To see this, on the left side of \figureref{figure:spread_predictions}, we plot the histogram of predictions made by the model which uses the entire set of available features (both individual and environmental features) for a specific school in the held-out test set. 
Even though the model has the ability to assign different predictions to different students, in practice, most students within this school have nearly identical predicted probabilities of graduation. 
Far from being an outlier, this particular histogram of within-schools predictions is representative of schools throughout the state. 
On the right side of \figureref{figure:spread_predictions}, we present the histogram of standard deviations of the within-school predictions made by this model across all schools in the test set.\footnote{That is, we group students in the held out test set according to their school IDs and compute the standard deviation of the predicted probabilities for these students. The histogram displays the frequencies of these standard deviations across all schools.} 

The main insight from this histogram is that the distribution of standard deviations is concentrated on small values. For the vast majority of schools, most students within the school receive almost exactly the same predicted probability of graduation.

This lack of within-school variation is in line with the initial experiment discussed earlier demonstrating the relative similarities in predictive performance between environmental and individual predictors. If there were large amounts of variation in the true probabilities of graduation within each school, a model that uses informative individual features would significantly outperform the environmental predictor that outputs a constant prediction.\footnote{
These conclusions are derived on the basis of the outputs on a predictor that was trained on finite amounts of data and used limited amounts of computation. Therefore, it is possible that the learning algorithm failed to pick up on existing variation in the true probabilities of graduation within specific schools. This is however, quite unlikely. Methods like gradient-boosted decision trees are widely believed to achieve Bayes’ risk for similar social science datasets where the number of data points far exceeds the number of features.}

\paragraph{Same School, Similar Individual Features.} This last observation provides further insight into the patterns observed within the DEWS system and the relative value of individual features.  We find that the very notion of an ``individual feature'' is somewhat fragile.  Due to the high levels of socioeconomic and racial segregation between public school districts, students within a particular district tend to have similar ``individual features''. If all students within the same school have the same individual features, then these features are largely meaningless for the sake of prediction. It becomes information-theoretically impossible to disambiguate different levels of dropout risk amongst students within the same school environment.

For example, one take away from the calibration plot in \figureref{figure:subgroupcalib} is that individual-level statistics such as race and socioeconomic status are so strongly correlated to the available environmental features, that it suffices to just know the environmental features in order to recover these individual features. Within the context of Wisconsin public schools, these two features are close to constant within a particular environment. This explains why the model that uses only environmental features can generated calibrated predictions on subgroups defined in terms of individual-level features.

Further experiments confirm this view. On average 75\% of the population in Wisconsin is non-White. However, a predictor that uses environmental features to predict whether a student is a person of color achieves a 0-1 error of .17, which is significantly  better than random prediction (random guessing gets .25). Similarly, 37.5\% of the students in our dataset qualify for free or reduced lunch. Yet, predicting free or reduced lunch status only using the environment achieves a 0-1 loss of .28. Random guessing would only achieve a misclassification error of .375 (lower numbers are better). 

These experimental results are conducted in the same fashion as the previous ones. We fit a predictor on the training set that only uses environmental features. However, in this case the target variable is the indicator for being a student of color, or (in a separate experiment) qualifying for free or reduced lunch. The accuracy of these models is evaluated on the held out test set.

Statistically speaking, even things like test-scores are environmentally determined. Standardized exams are explicitly designed to generate a Gaussian-like distribution of scores that distinguishes between high performing students and low performing students. 
However, most of the variance in this distribution comes from the fact that students in different schools score vary differently on the exam. Students within each school have similar scores. 

More formally, 80\% of schools have a within-school variance of math test scores that is smaller than the state-wide variance in test scores. If schools were composed of identical sub-populations of students, the within-district variance would be smaller than the state average only 50\% of the time.  Furthermore, this is not true just for individual math scores, if we consider the full vector of individual-features, 78.5\% of districts have lower variance than the state average.\footnote{That is, if we denote the vector of individual features by $x$, the variance of the random vector $\E[\norm{x - \E x}^2]$ is smaller when we take the expectation over $x$'s to be over students in a  specific school rather than the state as a whole.}

\end{document}